\documentclass[longauth]{aa}  
\usepackage[figuresright]{rotating}
\usepackage[authoryear]{natbib}
\bibpunct{(}{)}{;}{a}{}{,}

\usepackage{graphicx}
\usepackage{txfonts}
\newcommand{\etal}{\mbox{~et~al.}}
\newcommand{\xmm}{XMM-{\em Newton}}
\newcommand{\cosm}{XMM-{COSMOS}}
\newcommand{\flux}{erg cm$^{-2}$ s$^{-1}$}
\newcommand{\ecf}{cts s$^{-1}$ / 10$^{-11}$ erg cm$^{-2}$ s$^{-1}$}

\begin{document}
   \title{The \xmm~wide-field survey in the COSMOS field.}

   \subtitle{The  point-like X-ray source catalogue\thanks{Based on observations
   obtained with XMM--{\it Newton,} an ESA science mission with
instruments and contributions directly funded by ESA Member States and NASA;
also based on data collected at the Canada-France-Hawaii Telescope
operated by the National Research Council of Canada,  the Centre National de
la Recherche Scientifique de France and the University of Hawaii.}}

   \author{
	N. Cappelluti \inst{1,2}
          \and
          M. Brusa\inst{1,2}
	\and
	 G. Hasinger\inst{1}
	\and
	A. Comastri\inst{3}
	\and
	G. Zamorani\inst{3}
	\and
	A. Finoguenov\inst{1,2}
	\and
	R. Gilli\inst{3}
	\and
	S. Puccetti\inst{4}
	\and
	 T. Miyaji\inst{5} 
	\and
	M. Salvato\inst{6}
	\and
	C. Vignali\inst{7}
	\and
	T. Aldcroft\inst{8}
	\and
	H. B\"ohringer\inst{1}
	\and
	H. Brunner\inst{1}
	\and
	F. Civano\inst{8}
	\and
	M. Elvis\inst{8}
	\and
	F. Fiore\inst{9}
	\and
	A. Fruscione\inst{8}
	\and
	 R. E. Griffiths\inst{10}
	\and
	L. Guzzo\inst{11}
	\and
	A. Iovino\inst{11}
	\and
	A. M. Koekemoer\inst{12}
	\and
	V. Mainieri\inst{13}
	\and
	N.~Z. Scoville\inst{6}
	\and
	P. Shopbell\inst{6}
	J. Silverman\inst{14}
	\and
	C.~M. Urry\inst{15}
	 }

   \offprints{N. Cappelluti}

   \institute{ Max-Planck-Institut f\"ur Extraterrestrische Physik, Postfach 1312,
               85741, Garching bei M\"unchen, Germany
	  \and
         University of Maryland, Baltimore County, 1000 Hilltop Circle, Baltimore, MD 21250.
	\and
	 INAF-Osservatorio Astronomico di Bologna, via Ranzani 1, I-40127 Bologna, Italy
	\and
	 ASI Science Data Center, via Galileo Galilei, 00044 Frascati Italy.
	\and
	Instituto de Astronomia, Universidad Nacional Autonoma 
      de Mexico-Ensenada Km. 103 Carretera Tijuana-Ensenada, 22860 
  Ensenada, BC Mexico
	\and
	California Institute of Technology, 105-24 Robinson, 1200 East California Boulevard, Pasadena, CA 91125.
	\and
	 Dipartimento di Astronomia, Universit`a di Bologna, via Ranzani 1, I–40127 Bologna, Italy
	\and
	 Harvard-Smithsonian Center for Astrophysics, 60 Garden St, Cambridge, MA 02138
	\and
	 INAF-Osservatorio astronomico di Roma, Via Frascati 33, I-00044 Monteporzio Catone, Italy     
	\and
	 Department of Physics, Carnegie Mellon University, 5000
	Forbes Avenue, Pittsburgh, PA 15213
	\and 
	INAF-Osservatorio Astronomico di Brera - Via Brera 28, Milan, Italy
	\and
	Space Telescope Science Institute,3700 San Martin Drive, Baltimore, MD 21218
	\and
         ESO, Karl-Schwarschild-Strasse 2, D–85748 Garching, Germany
	\and	
	Institute of Astronomy, Department of Physics, Eidgen\"ossische Technische Hochschule, ETH Zurich,
	CH8093, Switzerland
	\and
	Department of Physics, Yale University, PO Box 208121, New Haven, CT 06520-8121, USA
	}

   \date{Received; accepted }
 
  \abstract{The COSMOS survey is a multiwavelength survey aimed to study the evolution
of galaxies, AGN and  large scale structures. Within this survey \cosm~a powerful tool
to detect AGN and galaxy clusters.
The \cosm~is a deep X-ray survey over the full 2 deg$^{2}$ of the COSMOS area. 
It consists of 55 \xmm~ pointings for a total exposure of$\sim$1.5 Ms with an average 
vignetting-corrected depth of 40 ks across the field of view and a sky coverage of 2.13 deg$^{2}$.} 
    { We present the catalogue of point-like   X-ray sources detected with the EPIC CCD cameras,
 	the logN-logS relations and the X-ray colour-colour diagrams.}
   {The analysis was performed using  the XMM-SAS data
     analysis package in the 0.5--2 keV, 2--10 keV and 5--10 keV energy bands.
 Source detection has been performed using a maximum likelihood technique especially
designed for raster scan surveys. The completeness of the catalogue  as well as logN-logS 
and source density maps have been calibrated  using Monte Carlo simulations.}
    {The catalogs contains a total of 1887 unique sources detected in at least one band
with likelihood parameter det\_ml$>$10. The survey, which shows
 unprecedented homogeneity,   has a flux limit of  $\sim$1.7$\times$10$^{-15}$ \flux,
 $\sim$9.3$\times$10$^{-15}$ \flux~ and $\sim$1.3$\times$10$^{-14}$ \flux~ over  90\% of the area (1.92 deg$^{2}$) 
 in the 0.5--2 keV,  2--10 keV and 5--10 keV energy band, respectively. 
Thanks to the rather homogeneous exposure over a large area, the derived logN-logS relations are very well 
determined over the flux range sampled by \cosm.
These relations have been compared with XRB synthesis models, which  reproduce  the observations with an
agreement of $\sim$10\% in the 5--10 keV and 2--10 keV
band, while in the 0.5--2 keV band the agreement is of the order of $\sim$20\%.
The hard X-ray colors confirmed that the majority of the extragalactic sources in a 
bright subsample are actually Type I or Type II AGN. 
About 20\% of the sources have a X-ray luminosity typical of AGN (L$_{X}>$10$^{42}$ erg/s) although they do not show 
any clear signature of nuclear activity in the optical
spectrum.}
    {}


   \maketitle
%

\section{Introduction}

The Cosmic evolution survey \citep[COSMOS,][]{scoville} with its 2 deg$^{2}$
of multiwavelength data is an exceptional laboratory to study  Active Galactic Nuclei (AGN), galaxies,  large scale structures 
of the Universe and their co-evolution. The survey uses multi-wavelength imaging and spectroscopy from X-ray to
radio wavelengths, including HST, Spitzer and GALEX imaging. 
The size of the survey has been chosen to sample 
large-scale structures with linear sizes of $\sim$50 Mpc $h^{-1}$ at z=1  with highly reduced
'cosmic' or sample variance.\\
During the AO3-AO4 and AO6 cycles, \xmm~surveyed   2.13 deg$^{2}$ of 
sky in the COSMOS field in the 0.5-10 keV energy band. The  
 total exposure was  $\sim$1.5 Ms split over 55 EPIC pointings.
The average resulting exposure  over the field of view is  $\sim$68 ks.
The central 0.9 deg$^{2}$ of the COSMOS field  also has been observed in X-rays with
 {\em Chandra} for a total of 1.8 Ms by  \citet{elvis08} (hereinafter C-COSMOS). \\
In this paper we present the X-ray pointlike source catalogue of the 1.5 Ms \cosm~survey together
with  the observation diary, data products,  logN-logS relations and colour-colour plots.
A subsample of the first year of \cosm~data has been presented in 
\citet{cap07} (hereafter Paper II) together with a detailed overview of the data analysis techniques. Here
we present data of all the observing cycles, with improved source positioning, higher
counting statistics and more  precise X-ray photometry.\\
Optical identifications of \cosm~sources, performed by 
taking advantage of the  precise  source positioning achieved with the complementary Chandra observations,
  will be presented in another paper \citep{brusa09}. \\
The  combination of the  moderately  deep flux limit and the wide  effective area 
 (flux limit of  $\sim$1.7$\times$10$^{-15}$ \flux~in the 
0.5-2 keV band over 1.92 deg$^{2}$) of the \cosm~made possible the compilation of a sample of  
sources with low influence of the so 
called sample or 'cosmic' variance.   Indeed, in paper II, assuming these survey parameters,
we estimated that  in \cosm~the fluctuations of the source density
due to cosmic variance are $<$5\%.
Furthermore, the tiling
of the observations was chosen to maximize the uniformity 
of the sensitivity over a large area of the field.\\
These particular characteristics, together with
 the multitude of  multiwavelength information available,
were  designed ad hoc to  study the large scale
 structures traced by X-ray emitting objects
like AGN and galaxy clusters and their co-evolution 
 \citep[see e.g.][]{cappi01,cap05,cap07a,bra07,koc08}.
 In addition these characteristics   make the survey    sensitive enough to study 
the evolution of super-massive black holes in the Universe up to high-z.
Considering the high throughput of \xmm~at high energies,
\cosm~will provide a valuable sample of absorbed sources 
to test X-ray background (XRB) synthesis model predictions. 
Moreover, to understand the nature of the XRB sources, 
 it is very important to have a detailed,
 cosmic variance free, measurement of the amplitude of the logN-logS relations in several 
energy bands. It is also
worth  noting that \cosm~ samples with 
good accuracy the flux range 
 where most of the XRB flux is produced (i.e. around S(2-10 keV)$\sim$10$^{-14}$ \flux). 
Therefore, among the  medium-deep X-ray surveys \citep{brandt},
  \cosm~ has the best combination of these characteristics
to achieve the goals mentioned above.\\
  The \cosm~survey, with its large area and counting statistics,  provides 
a large sample of bright sources where the hardness ratio can be measured 
with good precision.  Thanks also to the large amount of spectroscopic 
data in the field it is  possible to compare, in a reliable way,  
the optical properties with the X--ray properties derived from 
the hardness ratio analysis for large samples of sources. 
This is particularly important for AGN classification into absorbed (Type II) and unabsorbed  (Type I). 
In recent years it was realized (Szokoly et al. 2004)  that the classifications based on optical
spectroscopy may be affected by strong biases and AGN can be missed or not recognized as such. \\
The paper is organized as follows: in Section \ref{data} we present the  observations  and we
summarize the data reduction techniques;
in Section \ref{detection} we report on the source detection; 
in Section \ref{catalog} we present the pointlike source 
catalog; 
 in Section \ref{completeness} we quantify,
using Monte Carlo simulations, the completeness of the catalogue;
in Section \ref{logn} we present the logN-logS  relations;
in Section \ref{xrc} we give an overview of the source content
of the field using X-ray colour-colour diagrams  and the overall
 results are summarized in Section \ref{sum}. 
 Unless otherwise stated, errors are given at the
 1$\sigma$ level and we assume a $\Lambda$ dominated Universe 
 with H$_{0}$=70  km/s/Mpc, $\Omega_{m}$=0.3 and $\Omega_{\Lambda}$=0.7.

\begin{figure*}[!t]
\begin{center}
\resizebox{0.75\textwidth}{!}
{\includegraphics[scale=0.5]{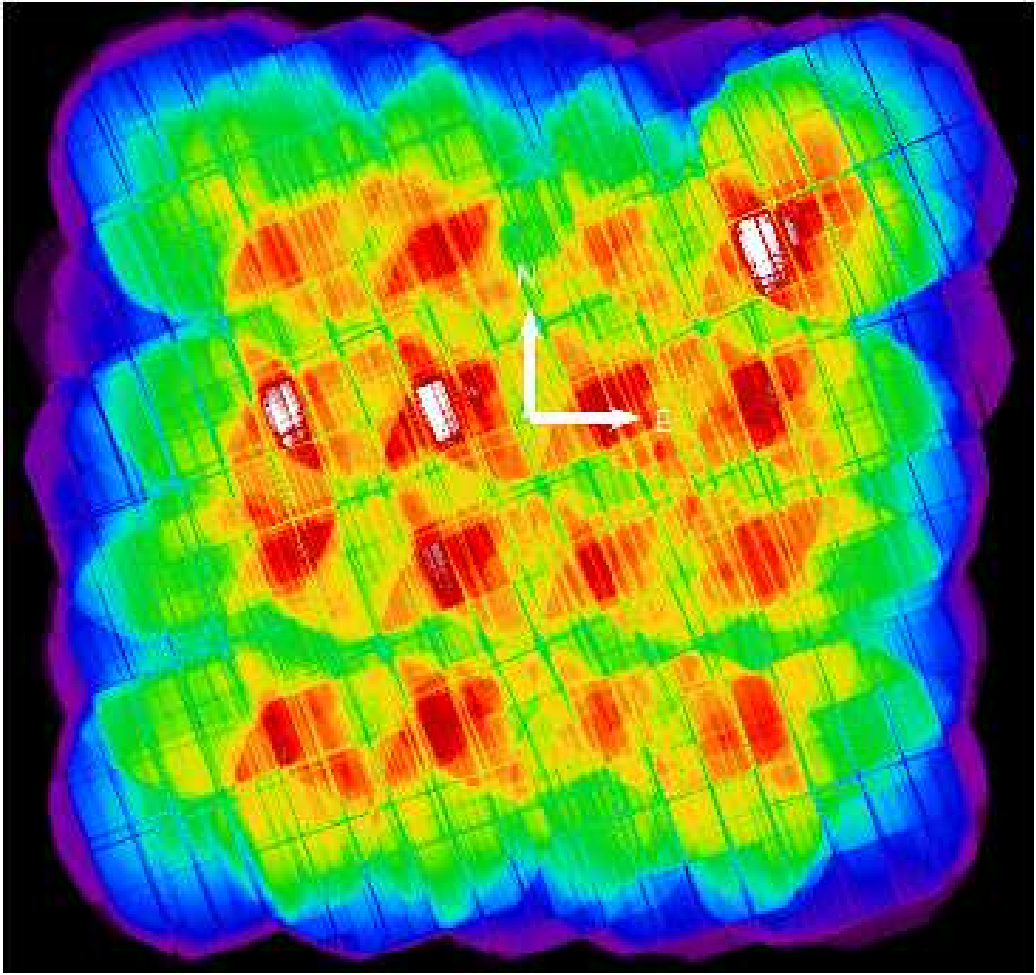}}
\caption{\label{fig:expmap}  Colour coded vignetting corrected 0.5--2 keV exposure map of the \cosm~survey. 
The maximum effective depth achieved on the field is $\sim$84 ks ($white$) and the mean exposure is  $\sim$68 ks ($green$).}
\end{center}
\end{figure*}
 \begin{figure*}[!t]
\begin{center}
\resizebox{0.75\textwidth}{!}
{\includegraphics[scale=0.5]{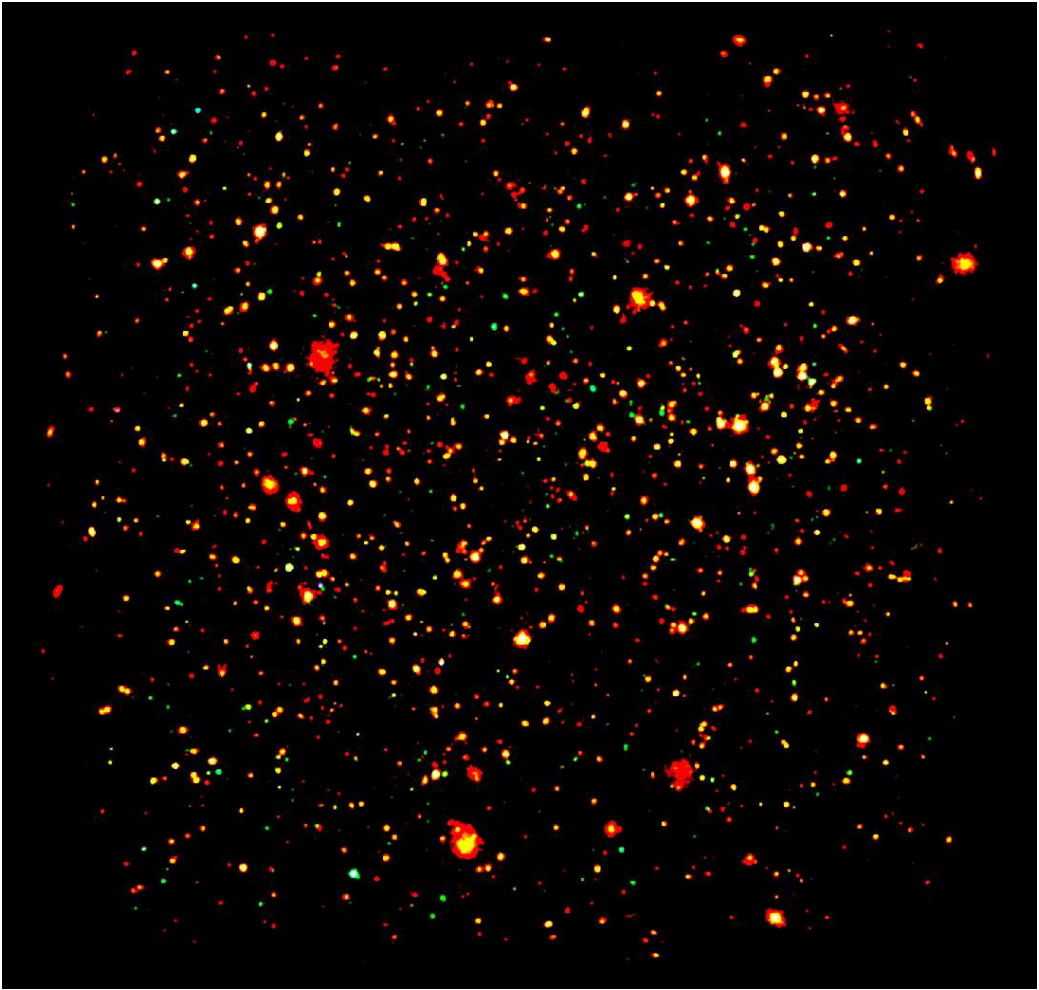}}
\caption{\label{fig:rgb} False colour
 X-ray image of the COSMOS field: red,green and blue colours represent the 0.5--2 keV, 2--4.5 keV and 4.5--10 keV 
energy bands, respectively. }
\end{center}
\end{figure*}

\section{\label{data} Observations and data reduction}

\begin{table*}[!t]
\begin{tiny}
\begin{center}
\begin{tabular}{ccccccccc}
\hline

ID & Revolution & OBS\_ID &  Date          & R.A. & DEC & EXPOSURE & GTI PN & GTI MOS\\
   & & & YYYY-MM-DDTHH:MM:SS & h m s & $^{\circ}~~\arcmin~~\arcsec$  & ks & ks & ks \\  	
\hline 
   1A$^{a}$ & 917 & 0203360101 & 2004-12-11T13:23:10 &  10  02 26.4 & 2 42 36.0 & 30.8 &  28.8 &  30.5 \\ 
   1B & 1090 & 0302350101 & 2005-11-21T20:57:04 &  10  02 26.4 & 2 43 36.0 & 19.9 &  14.2 &  17.8 \\ 
   2A$^{a}$ & 917 & 0203360201 & 2004-12-11T22:36:27 &  10  02 26.4 & 2 27 36.0 & 44.1 &  13.7 &  15.9 \\ 
   2B & 1190 & 0302350201 & 2006-06-08T15:28:17 &  10  02 30.4 & 2 27 36.0 & 19.9 &  12.5 &  16.1 \\ 
   3A$^{a}$ & 994 & 0203360301 & 2005-05-14T03:18:03 &  10  02 26.4 & 2 12 36.0 & 32.2 &  30.1 &  31.9 \\ 
   3B & 1083 & 0302350301 & 2005-11-07T08:41:36 &  10  02 26.4 & 2 13 36.0 &  5.3 &   1.6 &   4.3 \\ 
   3C & 1186 & 0302353101 & 2006-06-01T09:43:42 &  10  02 26.4 & 2 13 36.0 & 20.8 &  16.5 &  19.0 \\ 
   4A$^{a}$ & 907 & 0203360401 & 2004-11-21T05:12:10 &  10  02 26.4 & 1 57 36.0 & 30.8 &  25.6 &  29.1 \\ 
   4B & 1090 & 0302350401& 2005-11-22T03:07:03 &  10  02 30.4 & 1 57 36.0 & 25.1 &   7.3 &  11.6 \\ 
   5A$^{a}$ & 907 & 0203360501 & 2004-11-21T14:25:29 &  10  02 26.4 & 1 42 36.0 & 30.8 &  26.1 &  29.0 \\ 
   5B & 1089 & 0302350501 & 2005-11-19T16:32:05 &  10  02 26.4 & 1 43 36.0 & 19.9 &  17.9 &  19.7 \\ 
   6A$^{a}$ & 819 & 0203360601 & 2004-05-30T00:49:22 &  10  01 26.4 & 2 42 36.0 & 30.8 &  22.1 &  24.8 \\ 
   6B & 1186 & 0302350601 & 2006-06-01T03:33:43 &  10  01 22.4 & 2 42 36.0 & 19.9 &  15.4 &  18.8 \\ 
   7A$^{a}$ & 731 & 0203360701 & 2003-12-06T01:35:44 &  10  01 26.4 & 2 27 36.0 & 34.4 &  31.9 &  34.2 \\ 
   7B & 1091 & 0302350701 & 2005-11-23T05:06:10 &  10  01 26.4 & 2 28 36.0 & 19.9 &  17.8 &  19.1 \\ 
   8A$^{a}$ & 905 & 0203360801 & 2004-11-17T21:49:38 &  10  01 26.4 & 2 12 36.0 & 53.0 &  26.8 &  36.5 \\ 
   8B & 1092 & 0302350801 & 2005-11-25T19:44:36 &  10  01 22.4 & 2 12 36.0 & 19.9 &  17.6 &  19.4 \\ 
   9A$^{a}$ & 906 & 0203360901 & 2004-11-20T00:46:35 &  10  01 26.4 & 1 57 36.0 & 36.2 &  20.8 &  23.9 \\ 
   9B & 1095 & 0302350901 & 2005-12-02T02:52:30 &  10  01 26.4 & 1 58 36.0 & 24.4 &  11.0 &  11.8 \\ 
   9C & 1179 & 0302353001 & 2006-05-18T12:17:32 &  10  01 26.4 & 1 58 36.0 &  9.9 &   2.4 &   4.5 \\ 
  10A$^{a}$ & 907 & 0203361001 & 2004-11-21T23:38:52 &  10  01 26.4 & 1 42 36.0 & 45.5 &  12.9 &  17.2 \\ 
  10B & 1088 & 0302351001 & 2005-11-17T04:04:51 &  10  01 22.4 & 1 42 36.0 & 43.5 &  37.2 &  42.8 \\ 
  11A$^{a}$ & 912 & 0203361101 & 2004-12-01T23:23:41 &  10  00 26.4 & 2 42 36.0 & 44.2 &  19.5 &  22.8 \\ 
  11B & 1176 & 0302351101 & 2006-05-12T09:13:47 &  10  00 26.4 & 2 43 36.0 & 45.6 &  16.5 &  18.8 \\ 
  12A$^{a}$ & 732 & 0203361201 & 2003-12-08T18:19:32 &  10  00 26.4 & 2 27 36.0 & 34.9 &  25.1 &  26.6 \\ 
  12B  & 1091 & 0302351201 & 2005-11-23T11:16:10 &  10  00 30.4 & 2 27 36.0 & 19.9 &  13.9 &  15.8 \\ 
  13A$^{a}$ & 733 & 0203361301 & 2003-12-10T11:23:58 &  10  00 26.4 & 2 12 36.0 & 31.8 &  25.3 &  26.5 \\ 
  13B & 1091 & 0302351301 & 2005-11-23T17:26:09 &  10  00 26.4 & 2 13 36.0 & 19.9 &  18.0 &  19.2 \\ 
  14A$^{a}$ & 733 & 0203361401 & 2003-12-10T01:52:22 &  10  00 26.4 & 1 57 36.0 & 32.0 &  30.1 &  31.1 \\ 
  14B  & 1182 & 0302351401 & 2006-05-24T03:48:33 &  10  00 30.4 & 1 57 36.0 & 23.0 &  10.4 &  19.4 \\ 
  15A$^{a}$ & 906 & 0203361501 & 2004-11-19T15:33:15 &  10  00 26.4 & 1 42 36.0 & 30.9 &  20.2 &  26.9 \\ 
  15B & 1179 & 0302351501 & 2006-05-18T06:07:33 &  10  00 26.4 & 1 43 36.0 & 19.9 &  12.6 &  16.0 \\ 
  16A$^{a}$& 914 & 0203361601 & 2004-12-05T23:28:32 &  09 59 26.4 & 2 42 36.0 & 41.1 &  0.0 &   0.0 \\ 
  16B & 1093 & 0302351601 & 2005-11-27T17:59:55 &  09 59 22.4 & 2 42 36.0 & 57.3 &  28.2 &  36.6 \\ 
  17A$^{a}$ & 917 & 0203361701 & 2004-12-11T03:53:07 &  09 59 26.4 & 2 27 36.0 & 31.9 &  29.9 &  31.4 \\ 
  17B & 1179 & 0302351701 & 2006-05-17T23:57:32 &  09 59 26.4 & 2 28 36.0 & 19.9 &  17.7 &  19.6 \\ 
  18A$^{a}$ & 734 & 0203361801 & 2003-12-11T22:33:13 &  09 59 26.4 & 2 12 36.0 & 28.9 &  26.2 &  27.7 \\ 
  18B & 1179 & 0302351801 & 2006-05-17T17:47:33 &  09 59 22.4 & 2 12 36.0 & 19.9 &  16.8 &  18.6 \\ 
  19A$^{a}$ & 918 & 0203361901 & 2004-12-12T21:37:00 &  09 59 26.4 & 1 57 36.0 & 30.9 &  23.3 &  25.3 \\ 
  19B & 1178 & 0302351901 & 2006-05-15T23:34:59 &  09 59 26.4 & 1 58 36.0 & 19.9 &   9.9 &  17.9 \\ 
  20A$^{a}$& 994 & 0203362001& 2005-05-14T12:52:14 &  09 59 26.4 & 1 42 36.0 & 31.9 &   7.0 &   9.2 \\ 
  20B & 1178 & 0302352001 & 2006-05-15T17:24:58 &  09 59 22.4 & 1 42 36.0 & 19.9 &   4.9 &  16.6 \\ 
  20C$^{b}$& 1356 & 0501170101  & 2007-05-06T00:23:52 &  09 59 22.4 & 1 42 36.0 & 33.9 &  32.0 &  33.3 \\ 
  21A$^{a}$ & 916 & 0203362101 & 2004-12-09T07:16:01 &  09 58 26.4 & 2 42 36.0 & 62.6 &  60.3 &  61.7 \\ 
  22A$^{a}$ & 898 & 0203362201 & 2004-11-03T06:02:44 &  09 58 26.4 & 2 27 36.0 & 30.9 &  28.0 &  30.5 \\ 
  22B & 1176 & 0302352201 & 2006-05-11T19:47:08 &  09 58 30.4 & 2 27 36.0 & 21.9 &   7.0 &  10.8 \\ 
  23A$^{a}$ & 992 & 0203362301 & 2005-05-09T19:01:30 &  09 58 26.4 & 2 12 36.0 & 30.9 &   1.3 &  28.1 \\ 
  23B & 1176 & 0302352301 & 2006-05-12T02:30:29 &  09 58 26.4 & 2 13 36.0 & 21.9 &   4.3 &   7.4 \\ 
  23C$^{b}$ & 1362 & 0501170201& 2007-05-18T03:17:39 &  09 58 26.4 & 2 13 36.0 & 36.0 &  28.1 &  33.9 \\ 
  24A$^{a}$ & 992 & 0203362401 & 2005-05-10T04:14:50 &  09 58 26.4 & 1 57 36.0 & 30.9 &  17.4 &  23.0 \\ 
  24B & 1175 & 0302352401 & 2006-05-09T19:36:30 &  09 58 30.4 & 1 57 36.0 & 24.9 &   0.2 &  21.9 \\ 
  24C & 1190 & 0302353201 & 2006-06-09T01:36:12 &  09 58 30.4 & 1 57 36.0 & 19.3 &   9.7 &  14.6 \\ 
  25A$^{a}$ & 992& 0203362501 & 2005-05-10T13:28:11 &  09 58 26.4 & 2 42 36.0 & 31.9 &   0.0 &   0.0 \\ 
  25B & 1175 & 0302352501 & 2006-05-10T03:09:54 &  09 58 26.4 & 1 43 36.0 & 24.5 &  22.6 &  23.9 \\ 
  25C & 1190 & 0302353301 & 2006-06-09T07:36:11 &  09 58 26.4 & 1 43 36.0 & 18.9 &  11.6 &  14.3 \\ 
  \hline
\end{tabular}
\caption{\label{tab:logobs} The \xmm~ observation log of the \cosm~survey. From left to right: Field ID,revolution, OBS\_ID, observation start, 
right ascension, declination, duration of the exposure, Good Time Interval (GTI) for the PN and MOS camera, respectively.}
\end{center}
$^{a}$: Fields  observed in \xmm~AO3 presented by \citet{has07} and used for Paper II.\\
$^{b}$: Fields observed in \xmm~AO6.\\
\end{tiny}
\end{table*}
\begin{table*}[!t]
\begin{center}

\begin{tiny}
\begin{tabular}{ccccccc} 
\hline
\hline
 Band  & Total detections  & Single-Band detections & S$_{lim}$ & S$_{50\%}$ & S$_{90\%}$  & S$_{fa}$\\
       &                    &                  & \flux/10$^{-15}$ & \flux/10$^{-15}$ & \flux/10$^{-15}$ & \flux/10$^{-15}$\\
\hline
0.5--2 keV & 1621 & 771 & 0.50   & 1.00 &   1.70  & 3.00\\
2--10 keV  & 1111 & 237 & 2.50 &  5.60&   9.30    & 15.00    \\
5--10 keV  & 251  & 5   & 5.10  &  11.00 &   13.0 & 20.00    \\
\hline
\end{tabular}
\end{tiny}
 
\end{center}
\caption{\label{tab:sum} Summary of the total detections, single band detections, faintest flux limits,  flux limits at 50\%, 90\% of the 
total area  and  flux limits observable on 
the full area in the 0.5--2 keV, 2--10 keV and 5--10 keV energy bands, respectively.}
\end{table*}
The \cosm~survey  covers 2.13 deg$^{2}$ in the equatorial sky in a region bounded 
 by 9$^{h}$57.5$^{m}<\alpha<$10$^{h}$03.5$^{m}$ and 1$^{\circ}$27.5$\arcmin< \delta<$2$^{\circ}$57.5$\arcmin$. 
X-ray observations were performed during \xmm~AO3-AO4   from December 2003 to June 2006. 
The survey consists of a matrix of 5$\times$5 pointings shifted by 15$'$ with respect to each other. 
The matrix of pointings was observed in AO3  and repeated with a rigid shift of 1$'$ in AO4. The shift was applied to 
smooth sensitivity drops introduced by the CCD gaps. 
 In Table \ref{tab:logobs}  we present the log of the 55 \xmm~observations of the COSMOS field.\\
Because of charged particle flares,  two pointings were completely lost, 
namely 16A and 25A.  The lost times were compensated for by tuning the exposures in 
AO4. Additionally, two pointings  (i.e. field 20C and 23C)  were re-observed in \xmm~AO6
  (May 2007) for 32 ks each to compensate for time losses.
 At the time of writing  no further  observing campaigns of the COSMOS field are planned with \xmm.\\
 In Paper II we analyzed a  first sample of 23 fields observed with \xmm~during AO3
labeled in Table \ref{tab:logobs}. The total exposure was $\sim$504 ks after the cleaning of the   
particle background flares. The faintest sources in the field have a flux  of 
7$\times$10$^{-16}$ \flux, 4$\times$10$^{-15}$ \flux~and 9$\times$10$^{-15}$ \flux~in the 0.5--2 keV,
2--10 keV and 5--10 keV energy bands, respectively, while  a flux limit of  $\sim$1.7$\times$10$^{-15}$ \flux,
 $\sim$9.3$\times$10$^{-15}$ \flux~ and $\sim$1.3$\times$10$^{-14}$ \flux~ was
 achieved over  90\% of the area (1.92 deg$^{2}$) 
 in the 0.5--2 keV,  2--10 keV and 5--10 keV energy band, respectively.
 The preliminary catalogue based on those data consisted of 1390
independent sources and 1281, 784 and 186 source in the three bands, respectively.
We used that catalog to produce the first \cosm~logN-logS relations as  well as the first
study of the cosmic or sample variance in X-ray surveys. 
Paper II also contains a  detailed section on data analysis techniques, including
event cleaning, image processing, astrometry, source detection and Monte Carlo simulations.
In this  section we briefly summarize the analysis method;
we refer the reader to Paper II for a detailed description.\\ 
\xmm~was operated in imaging mode using the EPIC CCD cameras in full frame mode.  X-ray event files 
were searched for particle background flares and screened with the technique described in Paper II. 
In order to reduce the instrumental background, the energy channels between 1.45 keV and 1.54 keV were discarded
in both the MOS and PN data. To remove the strong Cu fluorescence features in the PN background
we also discarded the energy bands 7.2 keV-7.6 keV and 7.8 keV-8.2 keV. 
The total scheduled EPIC exposure  time was 1464 ks while, after the background cleaning the sum of the PN  
good time  intervals (GTI)  was $\sim$988 ks and 1207 ks for both  MOS1 and MOS2. \\
Due to the slow decrease of the solar activity from its  maximum in  
2000 to its minimum in 2007 \citep{hat99}, observations performed in AO3
and in the first part of A04 have a significantly higher  background level
than in the second part of AO4 and the two observations in AO6. 
 Event files were processed using the \xmm~Standard Analysis Software (SAS) version 6.7.0. After the 
removal of high background intervals we searched  for and removed hot/dead  columns and pixels.  
Images were created  in the 0.5--2 keV, 2--8 keV and 4.5--10 keV energy bands. In the same bands  
we created spectral weighted exposure  maps assuming a power-law model with photon spectral index $\Gamma$=2 in the 
 0.5--2 keV band and $\Gamma$=1.7 in the 2--8 keV and 4.5--10 keV bands.  \\
The  0.5--2 keV exposure map of the \cosm~survey is shown in 
Fig. \ref{fig:expmap}, while in Fig. \ref{fig:rgb} we show a false colour X-ray image of the entire field. \\
In order to compute background maps, we performed a  preliminary source detection using a sliding cell technique.
Using a threshold of 2.5$\sigma$ with the XMMSAS  software 'eboxdetect', we excised all the detected sources from all 
the images. The resulting images were  fitted with a double component model (a flat and a vignetted component)  to mimic
the particle  and the X-ray sky background.  \\
Astrometry corrections were  estimated as in Paper II by cross-correlating highly significant  (i.e. det\_ml$>$15,
 see Sect.  \ref{detection}) X-ray sources detected in each pointing, with the catalog  of galaxies detected in the I-band 
 by CFHT-MEGACAM \citep{mac07} and computing the most likely shift  using the XMM-SAS software "eposcorr".  
The   mean astrometric shift is similar to that reported in Paper II, being  $\Delta(\alpha)\sim$1.4" and $\Delta(\delta)\sim$0.2".

\section{ Source detection and source catalogue}

\subsection{ \label{detection} Source detection }
\begin{figure}[!t]
\begin{center}
\resizebox{\hsize}{!}
{\includegraphics{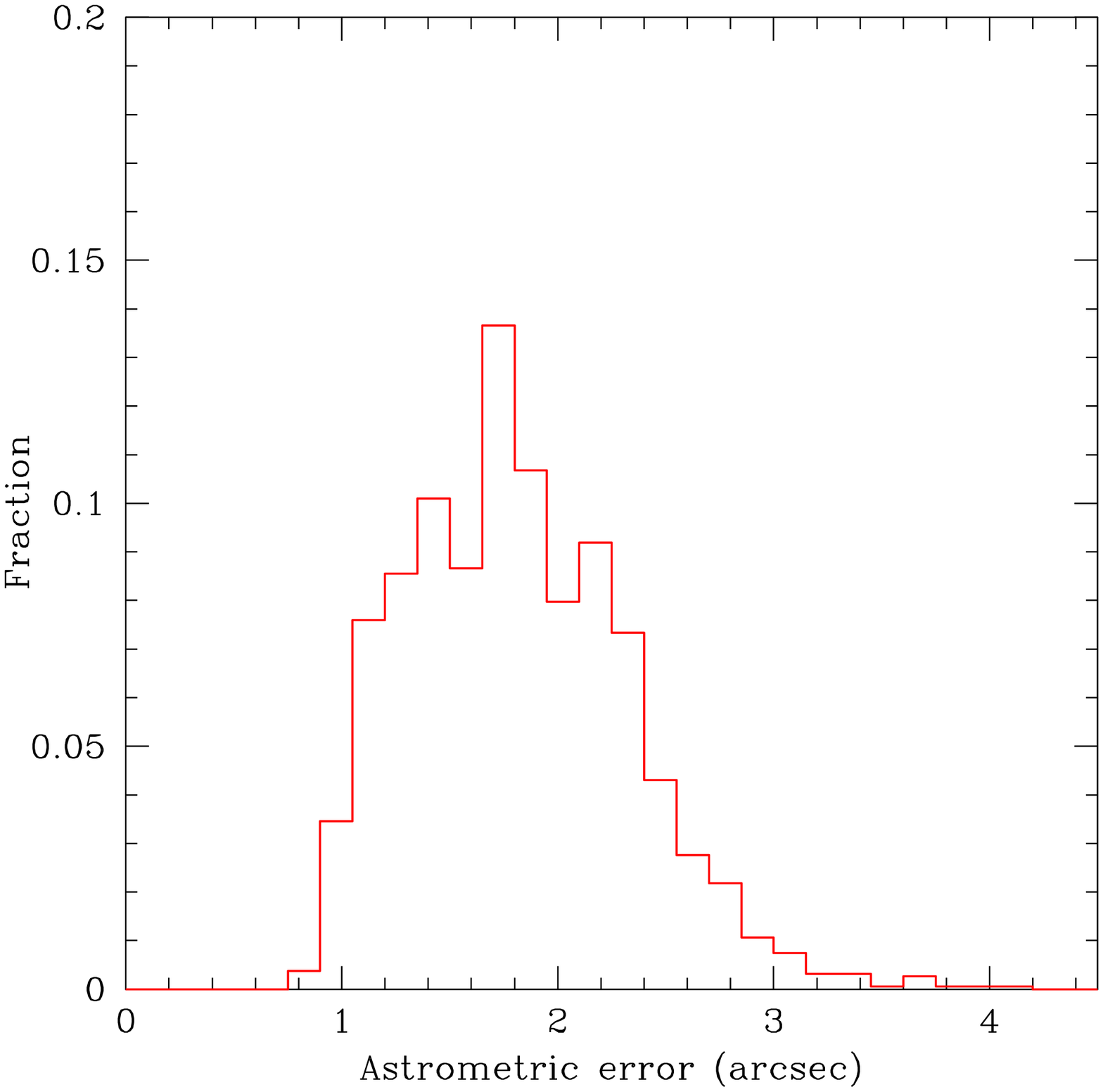}}
\caption{\label{fig:astro} The distribution of the positional uncertainties of the \cosm~ detections.   }
\end{center}
\end{figure}

We ran a two steps  source detection in three  energy bands, i.e the 0.5--2 keV, 2--8 keV and 4.5--10 keV. \\
By using the XMM-SAS tool "eboxdetect"  we first ran a  sliding cell detection to select  
 source candidates in the field. Differently from  Paper  II, we  used the 2--8 keV band in place of  the 2--4.5 keV band. 
This because a reanalysis of the AO3 data showed that the 2--8  keV band yields a better estimate
 of the 2--10 keV source counts (i.e. in the 2--4.5 keV band  we detected ~10\% fewer sources
than in the 2--8 keV band).  Moreover we determined that excluding the 8--10 keV events from the
analysis slightly enhanced the signal-to-noise ratio  of most the 2--10 keV sources. The 4.5--10 keV band has been fully exploited
in order to find the most absorbed sources.\\
If $P$ is the probability that a  Poissonian fluctuation of the background is detected as 
a spurious source, the likelihood of the detection is then defined as det\_ml=$-\ln(P)$. 
All the source candidates with det\_ml$<4$ were discarded.  
Making use of  the XMM-SAS tool "emldetect"  we then performed a maximum likelihood fit of each source candidate
 to a PSF model available in the \xmm~libraries.   
All the sources were also fitted with a convolution of a $\beta$-model cluster brightness profile \citep{cav} 
with the \xmm~PSF, to determine a possible extension in the
  detected signal. 
A source is classified  as  extended if the likelihood for the $\beta$-model fit 
exceeds  that of the pointlike case of 10 in det$\_ml$. 
The sources are fitted simultaneously in the  0.5--2 keV, 2--8 keV and 4.5--10 keV energy bands and  
the free parameters of the fit are position, flux and extension. \\
 Moreover, the calculation of the positional uncertainties in  each band also makes use  
 of the information in other bands and thus source 
positioning is extremely accurate in all the energy bands  ( see also discussion in Paper II).
 In Fig. \ref{fig:astro}  we show the distribution of the
statistical uncertainties on the source positions in arcsec as output by the  
emldetect task. 
The median statistical astrometric uncertainty, including also 
a systematic error of 0.75$\arcsec$ \citep[see][for a detailed discussion]{brusa09}, is
  1.77$\arcsec$\footnote{Similar results  also have been obtainedvia Monte Carlo simulations. }.
The reliability of the estimated source positions is confirmed by the distribution of the offset between 
X-ray sources  and optical counterparts. 
Count rates estimated in the 2--8 keV and 4.5--10 keV energy bands
were  extrapolated into 2--10 keV and 5--10 keV  fluxes, respectively.
In these bands we computed energy conversion factors (ECF) by assuming a power-law spectrum
 with spectral index $\Gamma$=1.7 and Galactic column density N$_{H}$=2.5$\times$10$^{20}$ cm$^{-2}$.  
In the 0.5--2 keV band, we directly converted the count-rate into fluxes assuming a spectral index $\Gamma$=2.0 
and Galactic column density N$_{H}$=2.5$\times$10$^{20}$ cm$^{-2}$.  The 
choice of these spectral indices is driven by the findings of \citet{mainieri07,mainieri08}. They
measured an average spectral index  $<\Gamma>\sim$1.7 for the \cosm~sources. 
In the soft band  we have chosen a   steeper index to  take into  account the 
contribution of the  soft excess.
Moreover these values of $\Gamma$ are widely used in the literature  \citep[see e.g.][]{has93,bal}  and therefore this choice
has also the scope of a better comparison with previous works, especially
when comparing the logN-logS relations. \\
The adopted ECFs\footnote{The ECF values  also take into account the energy channels discarded to 
decrease the background.} are 10.45 \ecf, 2.06 \ecf~ and 1.21 \ecf~ 
in the 0.5--2 keV, 2--8 keV and 4.5--10 keV energy bands, respectively.
All the sources with a maximum likelihood parameter det\_ml$>$10 in at least one band have been included
in the present catalog.   This threshold corresponds to a fraction of expected spurious
  sources of the order of 1.5\% in the 0.5-2 keV band and $\sim$0.5\%
in the other energy bands. Since in this work we used a more conservative  detection threshold than in Paper II (det\_ml$>$6),
the fraction of spurious sources has been significantly reduced.\\
 Significant detections have been achieved only in a subset of the energy bands. 
In the bands where the detection is not significant, we computed 1$\sigma$ 
upper limits of the counts using the prescriptions of \citet{nar06}.
Given M counts actually measured in a region of 30$\arcsec$ \footnote{We
checked that a 30$\arcsec$ aperture  gives the best 
agreement between aperture photometry  
and the  maximum likelihood PSF fitting technique.}  at the position of the source and
B background counts (estimated from our background maps), the 1$\sigma$ 
upper limit is defined as the number of counts $X$ that gives the probability
of observing M (or fewer) counts equal to the formal 68.3\%  Gaussian probability:
\begin{equation}
P(\leq M, X+B)=P_{Gauss}(68.3\%). 
\end{equation}
Assuming Poissonian statistics, this equation becomes:
\begin{equation}
\label{eq:ul}
P_{Gauss}=e^{-(X+B)}\sum_{i=0}^{M}\frac{(X+B)^{i}}{i!}.
\label{eq:uli} .
\end{equation}
By solving  Eq. \ref{eq:uli} iteratively in the case of $P_{Gauss}$=0.683,
we obtained the 1$\sigma$ upper limit X.
The upper limits were then converted into count-rates and fluxes
by diving by the exposure map and then applying the ECFs. 
We removed from the catalogue about 20 sources lying close to clear artifacts of the image 
(i.e. field and pointing boundaries or unremoved hot pixels).
With the method described above we selected a total of 1887 independent sources. 
Each source has been named with a unique ID number.
 1621 sources have been detected in the  0.5--2 keV energy band, while 
1111 and 251 sources are  detected  in the 2--10 keV and 5--10 keV band, respectively.
The number of sources with a significant detection in only one band is
771 for 0.5--2 keV band, 237 and 5 for the 2--10 keV and 5--10 keV bands, respectively.
The faintest sources in the field  have  fluxes  of 5.0$\times$10$^{-16}$ \flux, 
2.5$\times$10$^{-15}$ \flux and 5.1$\times$10$^{-15}$ \flux in the three  energy bands.\\
 A summary of the source detection
results is shown in Table \ref{tab:sum}.\\
Thanks to our PSF fitting technique we were able to detect 109  additional extended sources.
The catalog of the extended sources, together with a detailed and  more extensive analysis 
of their properties  will be presented in a forthcoming paper by \citet{fino08}.

\begin{sidewaystable*}[!h]
\begin{minipage}[t][180mm]{\textwidth}

\caption[]{\label{tab:cat} Extract of the source catalogue.}
\bigskip
\hspace{-2.9cm}
{\scriptsize
\begin{tabular}{lccccccccccccccccccccccc}
\hline
\hline
IAU Name & XID & Ra & Dec & Err & S & Cts & det\_ml & Bkg &Exp  & S & Cts$^{a}$ & det\_ml$^{a}$ & Bkg$^{a}$ & Exp$^{a}$  & S & Cts$^{b}$ & det\_ml$^{b}$ & Bkg$^{b}$ & Exp$^{b}$ \\  
 & & deg & deg & " & \flux/10$^{-14}$& & cts/pix & ks& &   \flux/10$^{-14}$ &  & &cts/pix &  ks &  \flux/10$^{-14}$ & &  &cts/pix & ks \\
 & & & & & 0.5--2 keV& & & &  & 2--10 keV & & & & & 5--10 keV& &  & \\ 
\hline
XMMU J100025.2+015850&     1&150.105148&  1.980817&0.91&  13.90 $\pm{ 0.18}$& 6585 $\pm{   83}$&21523.60& 0.80&45.45&  22.80 $\pm{ 0.70}$& 2210 $\pm{   49}$&4907.39& 0.95&46.96&12.20 $\pm{ 0.55}$&  625 $\pm{   27}$&900.04& 0.77&42.48\\
XMMU J095857.4+021313&     2&149.739190&  2.220533&0.93&  10.50 $\pm{ 0.18}$& 3875 $\pm{   66}$&10997.10& 1.10&35.34&  21.10 $\pm{ 0.72}$& 1588 $\pm{   39}$&3025.70& 1.79&36.55&10.70 $\pm{ 0.62}$&  406 $\pm{   23}$&475.59& 1.34&31.43\\
XMMU J095902.8+021906&     3&149.761543&  2.318492&0.90&  15.00 $\pm{ 0.17}$& 8621 $\pm{   99}$&24524.90& 2.00&54.87&  25.80 $\pm{ 0.73}$& 2924 $\pm{   61}$&5132.74& 2.91&55.05&14.20 $\pm{ 0.60}$&  783 $\pm{   33}$&874.38& 2.25&45.66\\
XMMU J095858.6+021458&     4&149.744181&  2.249476&0.97&   7.84 $\pm{ 0.17}$& 2561 $\pm{   55}$&6838.14& 1.12&31.25&  15.90 $\pm{ 0.79}$& 1066 $\pm{   39}$&1797.82& 1.82&32.59& 9.56 $\pm{ 0.69}$&  317 $\pm{   22}$&359.37& 1.37&27.54\\
XMMU J095918.8+020951&     5&149.828190&  2.164208&0.93&   7.28 $\pm{ 0.12}$& 4014 $\pm{   63}$&10945.70& 1.44&52.77&  14.80 $\pm{ 0.50}$& 1645 $\pm{   40}$&2852.41& 2.16&53.96& 7.54 $\pm{ 0.36}$&  433 $\pm{   20}$&457.84& 1.67&47.68\\
XMMU J100043.1+020636&     6&150.179776&  2.110154&0.96&   3.82 $\pm{ 0.08}$& 2966 $\pm{   59}$&6275.14& 2.05&74.30&   6.08 $\pm{ 0.33}$&  922 $\pm{   36}$&998.00& 2.94&73.67& 2.87 $\pm{ 0.28}$&  214 $\pm{   20}$&96.60& 2.34&61.92\\
XMMU J100205.1+023730&     7&150.521077&  2.625247&0.91&   9.66 $\pm{ 0.13}$& 5673 $\pm{   77}$&15117.20& 1.71&56.17&  18.90 $\pm{ 0.61}$& 2197 $\pm{   52}$&3494.22& 2.45&56.31&10.10 $\pm{ 0.52}$&  561 $\pm{   28}$&494.73& 1.95&45.94\\
XMMU J100012.9+023522&     8&150.053830&  2.589670&0.98&   5.78 $\pm{ 0.12}$& 2669 $\pm{   54}$&6651.50& 1.11&44.17&   5.40 $\pm{ 0.39}$&  490 $\pm{   26}$&535.94& 1.59&44.03& 1.94 $\pm{ 0.31}$&   88 $\pm{   14}$&32.45& 1.32&37.61\\
XMMU J095940.8+021938&     9&149.919828&  2.327475&1.00&   2.83 $\pm{ 0.07}$& 1905 $\pm{   47}$&3579.35& 1.45&64.42&   4.70 $\pm{ 0.30}$&  617 $\pm{   28}$&601.44& 1.98&63.77& 2.22 $\pm{ 0.27}$&  144 $\pm{   17}$&51.55& 1.67&54.12\\
XMMU J095939.0+021201&    10&149.912607&  2.200321&1.03&   2.72 $\pm{ 0.08}$& 1333 $\pm{   38}$&2763.96& 1.02&46.91&   2.21 $\pm{ 0.25}$&  220 $\pm{   18}$&195.79& 1.33&48.52& 1.09 $\pm{ 0.24}$&   55 $\pm{   12}$&14.53& 1.08&42.44\\
XMMU J100034.9+020234&    11&150.145317&  2.042825&1.04&   2.05 $\pm{ 0.06}$& 1307 $\pm{   38}$&2411.44& 1.50&60.94&   2.89 $\pm{ 0.20}$&  362 $\pm{   19}$&332.76& 2.04&61.00& 1.72 $\pm{ 0.26}$&  110 $\pm{   16}$&46.34& 1.65&53.07\\
XMMU J100049.9+020459&    12&150.207765&  2.083162&1.03&   2.22 $\pm{ 0.06}$& 1594 $\pm{   43}$&2915.84& 1.85&68.77&   2.40 $\pm{ 0.23}$&  340 $\pm{   23}$&207.47& 2.61&68.73& -0.61 $\pm{0.00}$&    0 $\pm{    0}$&-1&-1&  68.01 \\
XMMU J100002.2+021631&    13&150.009320&  2.275307&1.03&   2.11 $\pm{ 0.06}$& 1371 $\pm{   37}$&2288.04& 1.49&62.07&   2.90 $\pm{ 0.26}$&  369 $\pm{   24}$&287.04& 2.00&61.84& 1.10 $\pm{ 0.13}$&   69 $\pm{    8}$&15.30& 1.69&52.16\\
XMMU J100118.6+022739&    14&150.327292&  2.460850&1.03&   2.46 $\pm{ 0.08}$& 1024 $\pm{   32}$&2004.94& 0.76&39.84&   4.94 $\pm{ 0.32}$&  430 $\pm{   20}$&539.38& 0.88&42.28& 2.34 $\pm{ 0.23}$&  103 $\pm{   10}$&61.17& 0.71&36.60\\
XMMU J100159.8+022641&    15&150.499007&  2.444985&1.01&   3.11 $\pm{ 0.08}$& 1522 $\pm{   39}$&3305.96& 1.06&46.81&   5.80 $\pm{ 0.39}$&  567 $\pm{   28}$&765.08& 1.27&47.55& 3.39 $\pm{ 0.32}$&  170 $\pm{   15}$&123.67& 1.03&41.71\\
XMMU J100153.3+022437&    16&150.471935&  2.410327&1.05&   2.14 $\pm{ 0.07}$& 1114 $\pm{   36}$&2040.39& 1.26&49.72&   3.51 $\pm{ 0.30}$&  370 $\pm{   23}$&356.42& 1.57&51.20& 1.86 $\pm{ 0.30}$&   96 $\pm{   15}$&40.67& 1.27&43.01\\
XMMU J095924.5+015954&    17&149.852004&  1.998348&0.97&   4.37 $\pm{ 0.09}$& 2554 $\pm{   50}$&5958.98& 1.41&56.00&   6.10 $\pm{ 0.31}$&  705 $\pm{   26}$&814.66& 1.98&56.14& 2.85 $\pm{ 0.22}$&  170 $\pm{   13}$&110.64& 1.61&49.69\\
XMMU J100031.9+021811&    18&150.133028&  2.303236&1.04&   1.80 $\pm{ 0.06}$& 1098 $\pm{   34}$&1918.38& 1.29&58.49&   6.03 $\pm{ 0.34}$&  722 $\pm{   30}$&849.91& 1.70&58.11& 3.44 $\pm{ 0.29}$&  214 $\pm{   18}$&137.34& 1.40&51.68\\
XMMU J095958.5+021530&    19&149.993671&  2.258589&1.13&   1.49 $\pm{ 0.06}$&  748 $\pm{   30}$&821.62& 1.05&48.11&   2.82 $\pm{ 0.29}$&  280 $\pm{   21}$&232.51& 1.38&48.28& 1.10 $\pm{ 0.26}$&   55 $\pm{   12}$&11.47& 1.16&41.69\\
XMMU J100058.7+022556&    20&150.244637&  2.432327&1.09&   1.51 $\pm{ 0.05}$&  862 $\pm{   29}$&1475.05& 1.26&54.50&   2.30 $\pm{ 0.19}$&  257 $\pm{   16}$&206.81& 1.60&54.42& 1.28 $\pm{ 0.15}$&   73 $\pm{    8}$&25.79& 1.30&47.70\\
XMMU J100055.4+023441&    21&150.230771&  2.578229&1.04&   1.68 $\pm{ 0.05}$& 1281 $\pm{   39}$&1938.82& 1.79&73.07&   2.92 $\pm{ 0.24}$&  434 $\pm{   26}$&297.66& 2.49&72.27& 1.53 $\pm{ 0.23}$&  112 $\pm{   16}$&30.14& 2.05&60.81\\
XMMU J100046.7+020404&    22&150.194579&  2.067873&1.05&   1.75 $\pm{ 0.05}$& 1274 $\pm{   38}$&1994.07& 1.76&69.78&   1.75 $\pm{ 0.20}$&  250 $\pm{   21}$&146.61& 2.46&69.48& 0.67 $\pm{ 0.20}$&   58 $\pm{   14}$& 9.34& 1.99&58.26\\
XMMU J095909.6+021916&    23&149.789839&  2.321191&0.98&   3.69 $\pm{ 0.08}$& 2158 $\pm{   49}$&4172.35& 1.82&55.97&   5.53 $\pm{ 0.36}$&  644 $\pm{   30}$&619.50& 2.73&56.54& 2.76 $\pm{ 0.32}$&  156 $\pm{   18}$&70.86& 2.15&47.02\\
XMMU J100024.6+023149&    24&150.102567&  2.530383&1.08&   2.42 $\pm{ 0.08}$& 1131 $\pm{   35}$&2192.67& 1.04&44.77&   2.76 $\pm{ 0.29}$&  257 $\pm{   20}$&214.97& 1.40&45.33& 1.01 $\pm{ 0.24}$&   48 $\pm{   11}$&11.59& 1.16&40.19\\
XMMU J100024.5+020619&    25&150.102062&  2.105347&1.10&   1.38 $\pm{ 0.05}$&  737 $\pm{   28}$&1217.26& 1.02&51.18&   2.53 $\pm{ 0.25}$&  269 $\pm{   19}$&226.53& 1.30&51.63& 1.64 $\pm{ 0.24}$&   89 $\pm{   13}$&36.65& 1.08&45.07\\
XMMU J100135.9+024118&    26&150.399649&  2.688338&1.13&   1.23 $\pm{ 0.06}$&  583 $\pm{   26}$&910.32& 1.20&45.38&   3.08 $\pm{ 0.30}$&  296 $\pm{   21}$&279.27& 1.68&46.73& 1.01 $\pm{ 0.24}$&   49 $\pm{   11}$&11.84& 1.33&40.69\\
XMMU J100113.9+022547&    27&150.307994&  2.429963&1.09&   1.81 $\pm{ 0.06}$&  903 $\pm{   31}$&1434.77& 1.24&47.76&   2.79 $\pm{ 0.28}$&  283 $\pm{   20}$&226.68& 1.57&49.37& 1.37 $\pm{ 0.28}$&   68 $\pm{   13}$&18.30& 1.25&41.25\\
XMMU J100113.3+023607&    28&150.305619&  2.602049&1.07&   1.37 $\pm{ 0.05}$&  796 $\pm{   28}$&1186.44& 1.22&55.49&   3.26 $\pm{ 0.23}$&  377 $\pm{   19}$&308.79& 1.71&56.22& 1.41 $\pm{ 0.16}$&   82 $\pm{    9}$&23.35& 1.40&48.06\\
XMMU J095949.4+020140&    30&149.955983&  2.027961&1.11&   1.46 $\pm{ 0.05}$&  940 $\pm{   33}$&1454.97& 1.49&61.48&   1.56 $\pm{ 0.20}$&  196 $\pm{   19}$&114.14& 2.05&61.21& 0.61 $\pm{ 0.19}$&   43 $\pm{   11}$& 8.50& 1.69&51.75\\
XMMU J095946.9+022209&    31&149.945597&  2.369221&1.04&   1.74 $\pm{ 0.05}$& 1297 $\pm{   38}$&2096.72& 1.74&71.19&   2.79 $\pm{ 0.23}$&  404 $\pm{   25}$&294.20& 2.48&70.54& 1.27 $\pm{ 0.23}$&   89 $\pm{   16}$&17.05& 1.99&58.77\\
XMMU J095926.2+021529&    32&149.859221&  2.258144&1.14&   1.09 $\pm{ 0.04}$&  643 $\pm{   25}$&971.74& 1.18&56.25&   2.69 $\pm{ 0.26}$&  312 $\pm{   22}$&246.28& 1.55&56.32& 1.37 $\pm{ 0.15}$&   83 $\pm{    9}$&29.02& 1.30&50.65\\
XMMU J100114.3+022356&    33&150.309575&  2.399083&1.06&   1.56 $\pm{ 0.05}$&  971 $\pm{   33}$&1539.63& 1.50&59.43&   2.69 $\pm{ 0.25}$&  336 $\pm{   22}$&260.84& 1.94&60.53& 1.47 $\pm{ 0.24}$&   91 $\pm{   14}$&36.09& 1.53&51.45\\
XMMU J095958.5+021805&    34&149.993760&  2.301442&1.11&   1.22 $\pm{ 0.05}$&  885 $\pm{   33}$&1118.22& 1.63&69.35&   2.15 $\pm{ 0.22}$&  302 $\pm{   22}$&176.66& 2.22&68.25& 1.34 $\pm{ 0.23}$&   92 $\pm{   16}$&20.55& 1.88&57.16\\
XMMU J095928.3+022107&    35&149.868120&  2.351989&1.14&   1.27 $\pm{ 0.05}$&  719 $\pm{   28}$&1163.88& 1.13&54.14&   1.38 $\pm{ 0.21}$&  153 $\pm{   17}$&76.68& 1.46&54.06& -1.37 $\pm{ 0.00}$&    0 $\pm{    0}$&-1&-1&  53.50\\
XMMU J095940.1+022306&    37&149.916980&  2.385118&1.07&   1.44 $\pm{ 0.04}$& 1066 $\pm{   32}$&1651.10& 1.67&70.97&   2.41 $\pm{ 0.22}$&  347 $\pm{   23}$&263.57& 2.22&70.02& 1.20 $\pm{ 0.13}$&   88 $\pm{    9}$&26.36& 1.87&60.64\\
XMMU J100058.8+015359&    38&150.245126&  1.899753&1.05&   1.71 $\pm{ 0.05}$& 1199 $\pm{   37}$&1809.48& 2.00&66.95&   2.73 $\pm{ 0.25}$&  376 $\pm{   25}$&231.19& 2.85&66.99&  1.18 $\pm{ 0.00}$&   67 $\pm{   14}$& 9.24& 2.28&55.49\\
XMMU J100159.4+023934&    39&150.497618&  2.659684&1.10&   2.31 $\pm{ 0.07}$& 1077 $\pm{   32}$&1809.35& 1.25&44.62&   3.68 $\pm{ 0.35}$&  341 $\pm{   23}$&249.32& 1.78&45.03& 1.72 $\pm{ 0.20}$&   78 $\pm{    8}$&19.64& 1.39&37.69\\
XMMU J100114.8+020208&    40&150.311658&  2.035748&1.02&   2.14 $\pm{ 0.06}$& 1377 $\pm{   39}$&2421.11& 1.76&61.46&   3.83 $\pm{ 0.29}$&  488 $\pm{   27}$&408.01& 2.52&61.87& 2.16 $\pm{ 0.28}$&  137 $\pm{   17}$&53.29& 1.95&52.57\\
XMMU J100025.4+020734&    41&150.105662&  2.126228&1.17&   0.82 $\pm{ 0.04}$&  433 $\pm{   22}$&563.84& 0.98&50.90&    2.72 $\pm{ 0.27}$&  289 $\pm{   20}$&271.18& 1.24&51.60& 1.78 $\pm{ 0.26}$&   97 $\pm{   13}$&44.31& 1.03&45.36\\
XMMU J100202.8+022434&    42&150.511523&  2.409563&1.07&   1.70 $\pm{ 0.06}$& 1059 $\pm{   34}$&1706.14& 1.61&59.65&   2.46 $\pm{ 0.24}$&  303 $\pm{   21}$&198.07& 2.17&59.76& 1.10 $\pm{ 0.24}$&   66 $\pm{   14}$&11.44& 1.79&50.20\\
XMMU J095920.2+021831&    43&149.834184&  2.308670&1.15&   1.09 $\pm{ 0.05}$&  508 $\pm{   25}$&751.21& 0.81&44.70&    2.68 $\pm{ 0.28}$&  247 $\pm{   19}$&236.79& 1.06&44.86& 1.05 $\pm{ 0.24}$&   50 $\pm{   11}$&16.35& 0.88&39.66\\
XMMU J100051.5+021215&    44&150.214529&  2.204232&1.12&   1.20 $\pm{ 0.05}$&  670 $\pm{   27}$&1023.60& 1.13&53.60&   2.12 $\pm{ 0.23}$&  236 $\pm{   18}$&171.80& 1.50&54.30& 1.10 $\pm{ 0.23}$&   63 $\pm{   13}$&15.65& 1.16&48.06\\
XMMU J100120.6+022600&    45&150.335989&  2.433593&1.23&   0.54 $\pm{ 0.04}$&  321 $\pm{   21}$&333.23& 1.34&56.61&    2.09 $\pm{ 0.23}$&  251 $\pm{   20}$&214.50& 1.60&58.39& 1.38 $\pm{ 0.22}$&   86 $\pm{   13}$&34.75& 1.26&51.73\\
XMMU J100116.3+023606&    47&150.317898&  2.601891&1.14&   1.20 $\pm{ 0.05}$&  667 $\pm{   27}$&938.57& 1.19&53.30&    2.16 $\pm{ 0.24}$&  238 $\pm{   19}$&164.66& 1.55&53.37& 1.05 $\pm{ 0.22}$&   59 $\pm{   12}$&13.26& 1.29&47.04\\
XMMU J095905.2+021529&    48&149.771530&  2.258199&1.10&   1.37 $\pm{ 0.06}$&  690 $\pm{   28}$&978.65& 1.46&48.04&    3.42 $\pm{ 0.31}$&  342 $\pm{   23}$&297.98& 2.30&48.69& 2.43 $\pm{ 0.32}$&  122 $\pm{   16}$&51.30& 1.79&41.72\\
XMMU J100017.5+020011&    49&150.072853&  2.003149&1.18&   1.01 $\pm{ 0.05}$&  555 $\pm{   25}$&740.19& 1.05&52.75&    1.83 $\pm{ 0.23}$&  203 $\pm{   18}$&143.77& 1.39&54.18& 1.02 $\pm{ 0.22}$&   57 $\pm{   12}$&12.47& 1.12&47.07\\
XMMU J095934.1+021706&    50&149.891973&  2.285042&1.14&   0.93 $\pm{ 0.04}$&  598 $\pm{   26}$&756.33& 1.30&61.33&    2.30 $\pm{ 0.25}$&  290 $\pm{   22}$&206.48& 1.77&61.24& 1.58 $\pm{ 0.24}$&  101 $\pm{   15}$&31.08& 1.50&53.34\\
XMMU J100014.1+020053&    51&150.058859&  2.014971&1.15&   1.04 $\pm{ 0.04}$&  670 $\pm{   28}$&851.21& 1.40&61.47&    1.73 $\pm{ 0.21}$&  220 $\pm{   20}$&132.76& 1.88&61.91& 1.10 $\pm{ 0.21}$&   70 $\pm{   13}$&20.21& 1.54&53.24\\
XMMU J100016.3+015103&    52&150.067831&  1.850980&1.15&   1.14 $\pm{ 0.05}$&  581 $\pm{   25}$&829.17& 1.21&48.76&    1.91 $\pm{ 0.24}$&  194 $\pm{   18}$&99.87& 1.59&49.22& 1.74 $\pm{ 0.25}$&   50 $\pm{   12}$& 6.43& 1.29&42.46\\

\hline
\end{tabular}}

\vfill
$^a$: Estimated in the 2--8 keV band.\\
$^b$: Estimated in the 4.5--10 keV band.\\
\end{minipage}
\end{sidewaystable*}

\subsection{\label{catalog} Source catalogue}

\begin{figure}[h]
\begin{center}
\resizebox{\hsize}{!}
{\includegraphics{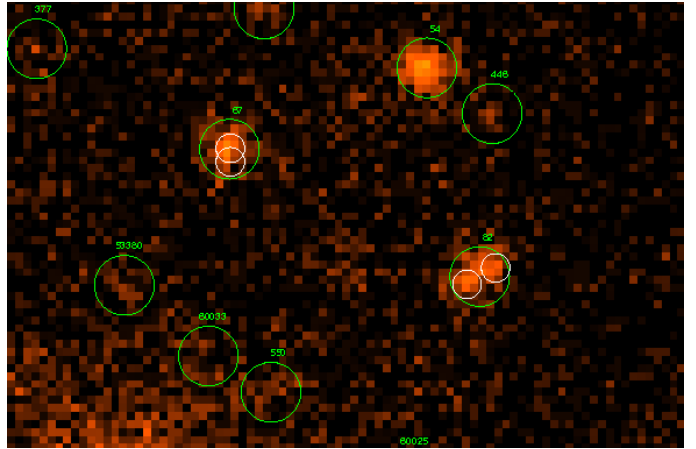}}
\caption{\label{fig:double} The region (6'$\times$3') containing the sources XID \#67 and \#82.  
Green circles  correspond to the 
\xmm~detections,  while white circles correspond to the Chandra detections.   }
\end{center}
\end{figure}

In Table. \ref{tab:cat}  we show, as an example, the first 50 entries of the catalogue as they appear on-line.
The table is structured as follows: \\
$Column~1$: IAU Name, $Column~2$: XID, $Column~3$: $\alpha$ (deg), $Column~4$: $\delta$ (deg),  $Column~5$: Positional error (arcsec),
$Column~6$: 0.5--2 keV flux  (\flux/10$^{-14}$), $Column~7$: 0.5--2 keV net counts , $Column~8$: 0.5--2 keV
 likelihood parameter det\_ml, $Column~9$: 0.5--2 keV background counts (cts/pix)\footnote{The pixel scale is  4``/pix},
 $Column~10$: 0.5--2 keV  vignetting corrected exposure (ks), $Column~11$: 2--10 keV 
flux (\flux/10$^{-14}$), $Column~12$: 2--8 keV net counts , $Column~13$: 2--8 keV likelihood 
parameter det\_ml, $Column~14$: 2--8  keV background counts (cts/pix), $Column~15$: 2--8 keV vignetting corrected  exposure (ks), 
$Column~16$:  5--10 keV flux, 
(\flux/10$^{-14}$), $Column~17$:  4.5--10 keV net counts, $Column~18$: 4.5--10 keV likelihood parameter det\_ml, 
 $Column~19$:  4.5--10  keV background counts,  $Column~20$: 5--10 keV vignetting corrected  exposure (ks).\\
The  interactive and machine readable full version of the catalog can downloaded at
\verb+http://irsa.ipac.caltech.edu/data/COSMOS/+. 
For sources with no significant
detection in a band, 
we list  upper limits with  negative values of  flux. In this case,
we also quote a value of cts=0, det\_ml=-1  and Bkg=-1 (background)  
in the band where the detection is not significant. \\
The flux  errors are the statistical uncertainties estimated
from the maximum likelihood and do not include uncertainties
introduced by the choice of the spectral model to estimate the
flux. We determined that by varying by $\Delta\Gamma$=0.3 the spectral index 
 assumed in computing the fluxes, 
the resulting variation of the flux estimate is of the order of 2\%, 9\%,and 4\% 
in the three bands.\\
 The Chandra coverage of the inner area of the XMM-COSMOS field \citep{elvis08}
offers a unique possibility to investigate the effect of source
confusion in our catalog.\\
The Chandra field covers about half of the XMM-COSMOS field.
Of the 1887 XMM-sources with   det\_ml$>$10, 946 (50.1\%) have been observed by
Chandra with an exposure longer than 30 ks, and  876 of them are present in
the C-COSMOS point-like source catalog \citep{elvis08}. 
Twenty-four of the 876 XMM  pointlike   sources with Chandra coverage (2.7\%)
are actually {\it resolved} into two different Chandra sources, which
lie between 2 and 10 arcsec from each other and have been blurred by
the XMM large PSF.
We then used the Chandra source counterpart positions of these 24
"blended sources" (48 different positions) as the input catalog for emldetect
and we fitted these sources keeping the position parameter fixed at the Chandra value.
As a result only 2/24 \cosm~sources have been  deblended into 4  \xmm~sources,
 namely XID \#67 and XID \#82, while the remaining 22 sources have been detected again as a single
\cosm~source with properties consistent with those
presented in the catalogue\footnote{We kept these sources as single entries in the catalog for self 
consistency with our statistical analysis}. Therefore we can conclude that our sample contains  $<$2.7\% of the sources
which could be resolved into two sources at the Chandra-COSMOS flux limit. 
In Fig. \ref{fig:double} we show the \xmm~image of a region containing 
the two deblended sources, which by chance are  close to each other. 
 \begin{figure*}[!t]
\begin{center}
\includegraphics[scale=0.45]{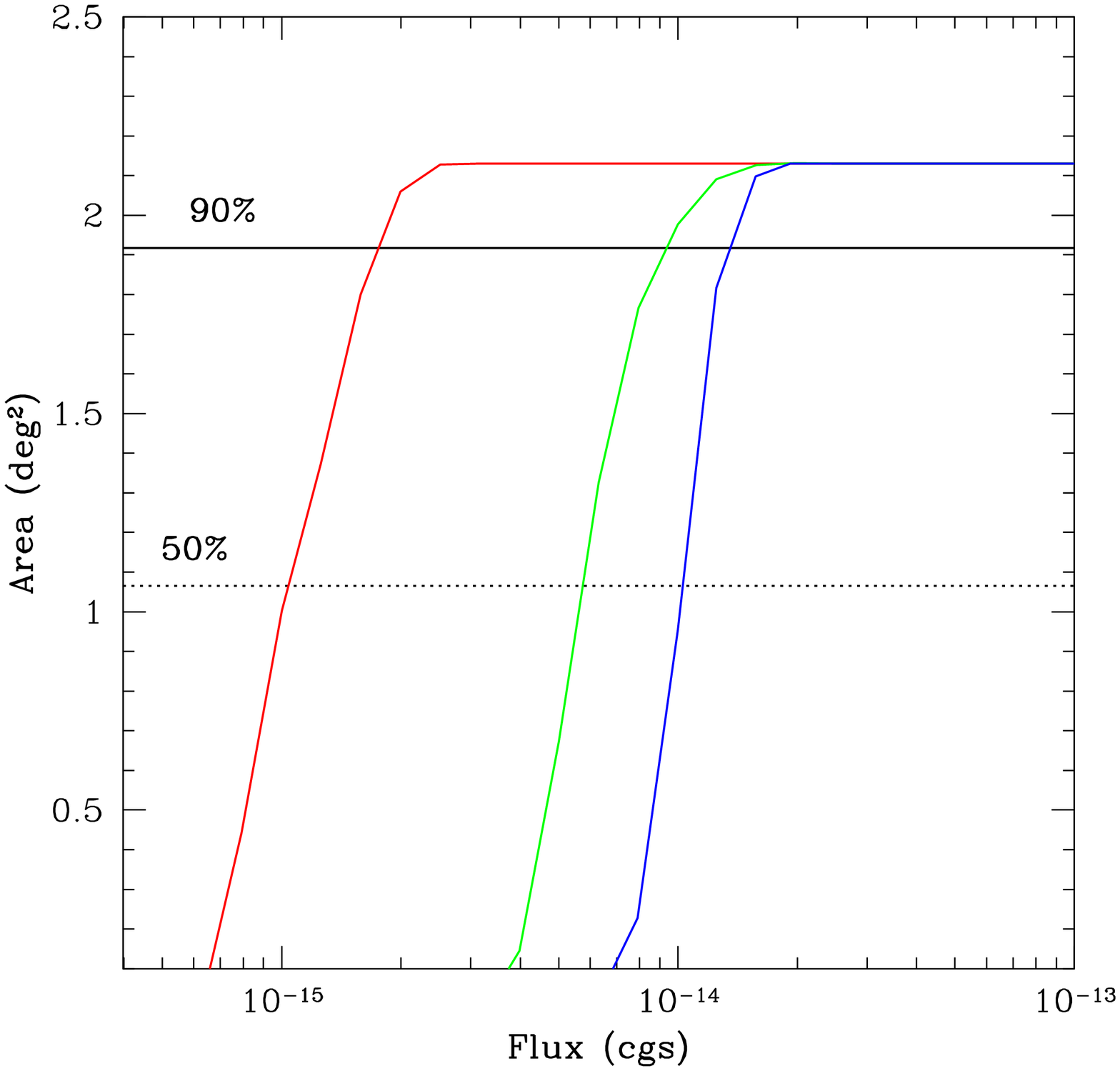}
\includegraphics[scale=0.45]{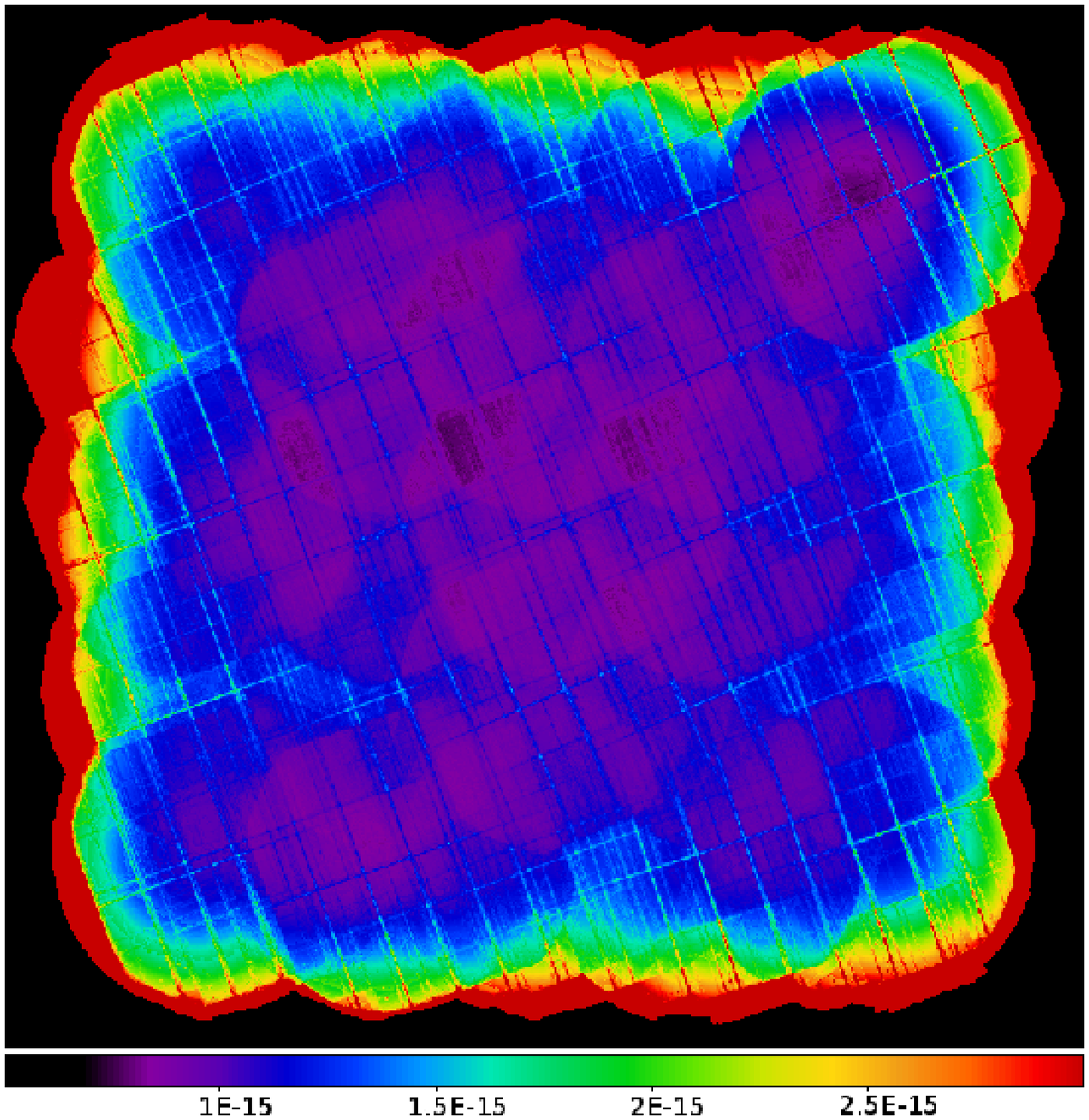}
\caption{\label{fig:skycov} $Left~Panel:$ The sky coverage versus 
flux for the 0.5--2 keV, 2--10 keV and 5--10 keV bands,
represented by  red, green and blue solid lines.
 The horizontal solid and dashed lines show the 90\% and 50\% completeness levels.
$Right~Panel:$ The 0.5--2 keV sensitivity map of \cosm~in  \flux. The map is plotted in colour
 coded scale from 5$\times$10$^{-16}$\flux~($magenta$) to 3$\times$10$^{-15}$\flux~($red$). }
\end{center}
\end{figure*}
 \subsection{\label{completeness} Monte Carlo simulations, sky coverage and sensitivity maps}
In order to estimate the sky coverage  of our survey, we  performed Monte Carlo simulations
as described in Paper II. The precision of the photometry as well as  positional accuracy were also discussed 
in Paper II.  
Here we give an overview of the procedure adopted for the production of random X-ray sky images and their analysis.\\
 Twenty series of 55 \xmm~  images   were created with the same pattern, exposure maps and background levels as the real data.
We produced 20 random input source catalogs with sources randomly placed in the field of view and fluxes distributed
 according to the AGN logN-logS distributions  predicted by the \citet{gilli07}  XRB synthesis model.	
 The input fluxes were  converted into count-rates by folding through the response matrices
the same spectral model assumed to compute fluxes and to weight exposure maps.  
The counts of the sources were  then convolved with  \xmm~PSF  templates available in the  \xmm~ calibration database and 
reproduced on the detector.
  We then applied to the simulated fields the same 
source detection procedure used in  the real data producing 20 independent output catalogs.  \\
 The sky coverage is then obtained by dividing the number of detected sources at each flux by the number of input sources
 and rescaling for a total area of 2.13 deg$^{2}$. 
    By using as a model the \citet{gilli07} logN-logS, it 
is possible that the simulated logN-logS could be slightly different from the real logN-logS. This could
introduce some biases in the estimation of the effective area. However, \citet{sch}
showed that the effect of a different slope of the logN-logS is negligible when the threshold of the source detection
is higher than 3-4$\sigma$.\\
 The  sky coverage in the three energy bands under investigation is plotted in the left panel of Fig. \ref{fig:skycov}
 As a result of the simulations, we obtained that  90\%  of the survey area is sensitive to  flux 
limits of $\sim$1.7$\times$10$^{-15}$ \flux, $\sim$9.3$\times$10$^{-15}$ \flux~ and $\sim$1.3$\times$10$^{-14}$ \flux~ 
in the three energy bands, respectively.
Additionally, we determined that the survey is
sensitive, over the full field of view (i.e. 2.13 deg$^{2}$), to  fluxes of $\sim$3.0$\times$10$^{-15}$ \flux,
 $\sim$1.5$\times$10$^{-14}$ \flux~ and $\sim$2.0$\times$10$^{-14}$ \flux~, respectively.
 As mentioned above, the fluxes of the input spectrum are converted into count-rates by assuming 
a single spectrum for all the sources. This could in principle bias the estimates of the 
sensitivity limits. In order to test the effect of a variation of the
mean spectral index in the estimate of the sky coverage,
 we changed the spectral indices by $\Delta\Gamma=\pm{0.3}$.
In this way the estimate of the flux limit changed by $<$2\% in the soft band, 
while 
this variation was of the order of 9\% and 4\% in the 2--10 keV and 5--10 keV band. 
The sky coverage is thus almost insensitive
to a change of the spectral shape in the 0.5--2 keV and   in the 5--10 keV band. In the 2--10  keV
we estimated that a 9\% uncertainty in the flux limit could introduce an overall
uncertainty of $\sim$5\% in the  logN-logS. Such an uncertainty is however smaller
than the typical  uncertainty on the source counts.   \\
In order to map the  sensitivity across the field of view
 we  produced sensitivity maps of the \cosm~survey in all the energy bands by reversing our source detection analysis. 
By using our estimated  background maps and exposure maps we evaluated, according to the Poisson statistic, the minimum number
of counts necessary to have  a detection with $det\_ml>$10. The number of counts have been evaluated 
in cells of 3$\times$3 pixels and corrected for the fraction of the PSF  falling out of the cell. The resulting count-limit maps have been divided by
the exposure maps and converted into  flux limit maps using the ECF. \\
 As an example, the resulting 0.5--2 keV band sensitivity map is plotted in colour scale in the $right~panel$ of
Fig. \ref{fig:skycov}. 
The map is in excellent agreement with the sky coverage plot obtained via Monte Carlo simulations.
 As one  can notice,  almost all the central area ($\sim$1.8 deg$^{2}$) 
has a quite homogeneous flux limit  $\sim$1.7$\times$10$^{-15}$ \flux.
 The northern central part of the field shows an area of $\sim$0.5 deg$^{2}$, 
 having a flux limit of the order of $\sim$8$\times$10$^{-16}$ \flux.
 It is worth  noting that the deepest part of
the field is located in the northeastern part of the field and the flux limit 
of $\sim$5.0$\times$10$^{-16}$ \flux~in the sensitivity maps is  in agreement with 
the predictions of the Monte Carlo simulations and the output of the source detection.
\begin{table*}[!t]

\caption[]{\label{tab:logn}  Source number counts.}
\begin{tabular}{ccccccc}
\hline
Log(S)  & N($>$S) & Area & N($>$S) & Area & N($>$S) & Area \\
\hline
\hline
\flux~ &  deg$^{-2}$ & deg$^{2}$ &   deg$^{-2}$ & deg$^{2}$ &   deg$^{-2}$ & deg$^{2}$ \\
  &  0.5--2 keV&   &    2--10 keV &  & 5--10 keV & \\
\hline
-12.8 &     1.51 $ \pm{   0.81}$ &   2.13  &     3.86 $ \pm{   1.33}$ &   2.13 &     1.04 $ \pm{   0.66}$ &   2.13 \\
-12.9 &     2.45 $ \pm{   1.05}$ &   2.13  &     6.20 $ \pm{   1.69}$ &   2.13 &     1.98 $ \pm{   0.94}$ &   2.13 \\
-13.0 &     3.39 $ \pm{   1.24}$ &   2.13  &     8.55 $ \pm{   1.99}$ &   2.13 &     2.92 $ \pm{   1.15}$ &   2.13 \\
-13.1 &     5.26 $ \pm{   1.56}$ &   2.13  &    11.84 $ \pm{   2.35}$ &   2.13 &     4.79 $ \pm{   1.48}$ &   2.13 \\
-13.2 &     8.08 $ \pm{   1.94}$ &   2.13  &    13.25 $ \pm{   2.48}$ &   2.13 &     7.14 $ \pm{   1.82}$ &   2.13 \\
-13.3 &    10.43 $ \pm{   2.20}$ &   2.13  &    24.51 $ \pm{   3.39}$ &   2.13 &     8.55 $ \pm{   1.99}$ &   2.13 \\
-13.4 &    14.65 $ \pm{   2.61}$ &   2.13  &    32.96 $ \pm{   3.93}$ &   2.13 &    11.84 $ \pm{   2.35}$ &   2.13 \\
-13.5 &    19.35 $ \pm{   3.01}$ &   2.13  &    50.33 $ \pm{   4.86}$ &   2.13 &    17.94 $ \pm{   2.89}$ &   2.13 \\
-13.6 &    25.92 $ \pm{   3.48}$ &   2.13  &    77.56 $ \pm{   6.03}$ &   2.13 &    24.98 $ \pm{   3.42}$ &   2.13 \\
-13.7 &    40.48 $ \pm{   4.35}$ &   2.13  &   108.07 $ \pm{   7.12}$ &   2.13 &    39.07 $ \pm{   4.28}$ &   2.13 \\
-13.8 &    56.44 $ \pm{   5.14}$ &   2.13  &   150.33 $ \pm{   8.40}$ &   2.13 &    55.57 $ \pm{   5.11}$ &   2.10 \\
-13.9 &    76.63 $ \pm{   5.99}$ &   2.13  &   208.52 $ \pm{   9.90}$ &   2.09 &    84.72 $ \pm{   6.36}$ &   1.82 \\
-14.0 &   102.45 $ \pm{   6.93}$ &   2.13  &   277.20 $ \pm{  11.48}$ &   1.98 &   123.75 $ \pm{   8.26}$ &   0.95 \\
-14.1 &   131.09 $ \pm{   7.84}$ &   2.13  &   361.50 $ \pm{  13.31}$ &   1.77 &   166.26 $ \pm{  13.33}$ &   0.23 \\
-14.2 &   166.77 $ \pm{   8.85}$ &   2.13  &   491.38 $ \pm{  16.27}$ &   1.33 &   212.88 $ \pm{  43.57}$ &   0.02 \\
-14.3 &   217.00 $ \pm{  10.09}$ &   2.13  &   620.89 $ \pm{  19.91}$ &   0.67 & & \\                                      
-14.4 &   273.34 $ \pm{  11.33}$ &   2.13  &   766.57 $ \pm{  29.63}$ &   0.14 & & \\                                       
-14.5 &   324.98 $ \pm{  12.35}$ &   2.13  &   984.00 $ \pm{ 106.76}$ &   0.01 & & \\                                       
-14.6 &   398.23 $ \pm{  13.67}$ &   2.13  & & & & \\                                                                          
-14.7 &   480.54 $ \pm{  15.04}$ &   2.06   & & & & \\                                                                           
-14.8 &   581.69 $ \pm{  16.68}$ &   1.80    & & & & \\                                                                          
-14.9 &   713.44 $ \pm{  19.05}$ &   1.37     & & & & \\                                                                         
-15.0 &   842.39 $ \pm{  21.69}$ &   1.00      & & & & \\                                                                        
-15.1 &   930.12 $ \pm{  24.30}$ &   0.44      & & & & \\                                                                        
-15.2 &  1027.83 $ \pm{  35.07}$ &   0.08       & & & & \\                                                                       
-15.3 &  1201.41 $ \pm{ 177.09}$ &   0.01      & & & & \\                                                                        
\hline

\end{tabular}
\end{table*}
\section{\label{logn}logN-logS relations}
\begin{figure}[!h]
\begin{center}
\resizebox{\hsize}{!}
{\includegraphics{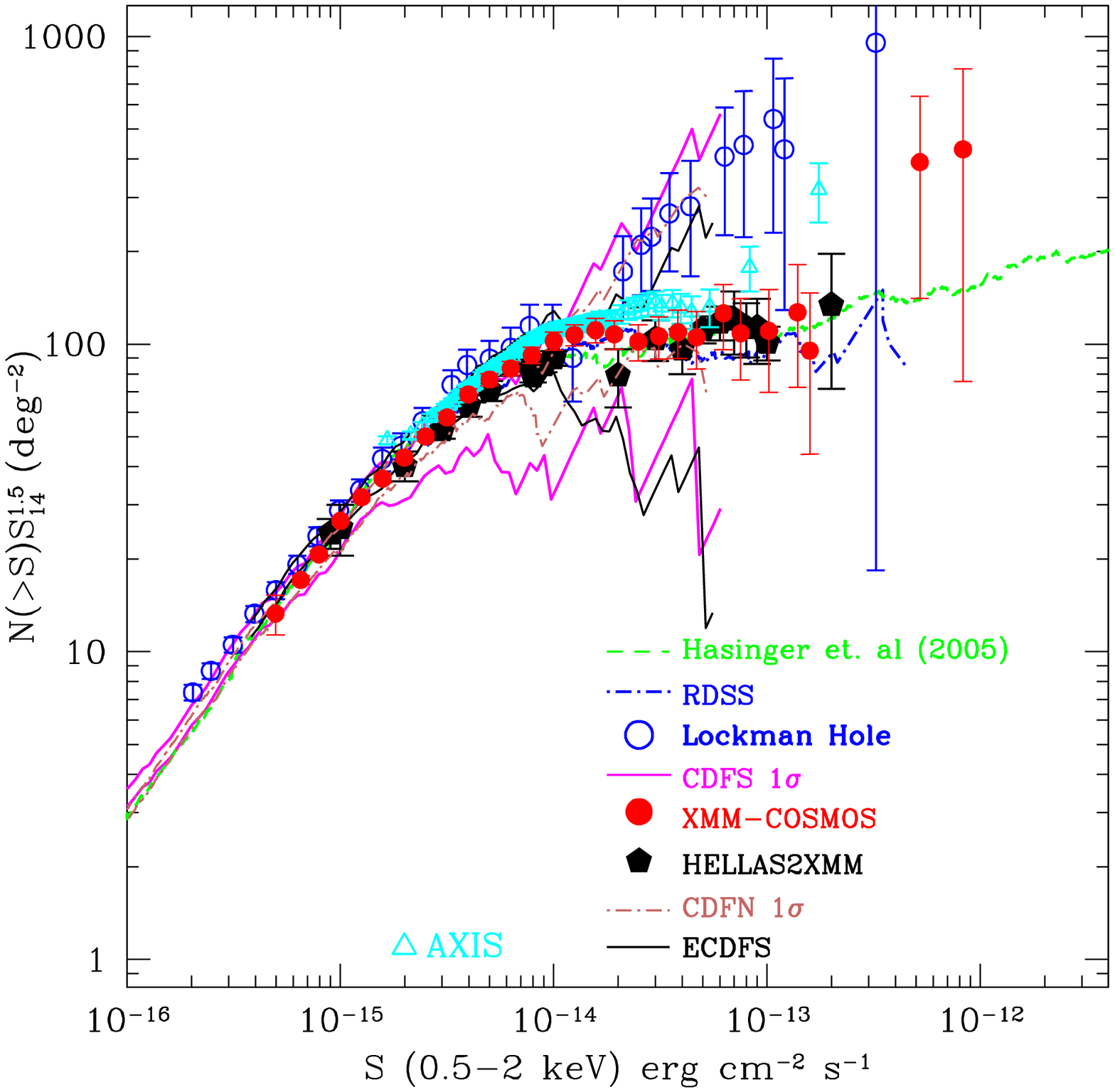}}
\caption{\label{fig:logn1}  The 0.5-2 keV logN-logS of the XMM-COSMOS ($red~dots$) sources compared with the ROSAT medium 
sensitivity survey \citep[][$blue~dot~dashed~line$]{has93},  combined ROSAT, \xmm ,  {\it Chandra} sources \citep[][$green~dashed~line$]{has05}, 
the 2Ms CDFS   \cite[1$\sigma$ error tie,][$magenta~continuous~line$]{luo}, the 2 Ms CDFN  \cite[1$\sigma$ error tie,][$pink~dot-dashed~line$]{bau04}, 
the \xmm ~ Lockman hole ~\cite[][$blue~circles$]{brunner08}, the AXIS \cite[][$cyan~triangles$]{carrera07}, the HELLAS2XMM \cite[][$black~pentagons$]{bal}   
and the extended CDFS \cite[1$\sigma$ error tie,][$black~continuous~line$]{leh05}  surveys. 
The source number counts are plotted multiplied by (S/10$^{14}$)$^{1.5}$ in order to highlight the  deviations from the Euclidean behavior.}
\end{center}
\end{figure}

\begin{figure}[!h]
\begin{center}
\resizebox{\hsize}{!}
{\includegraphics{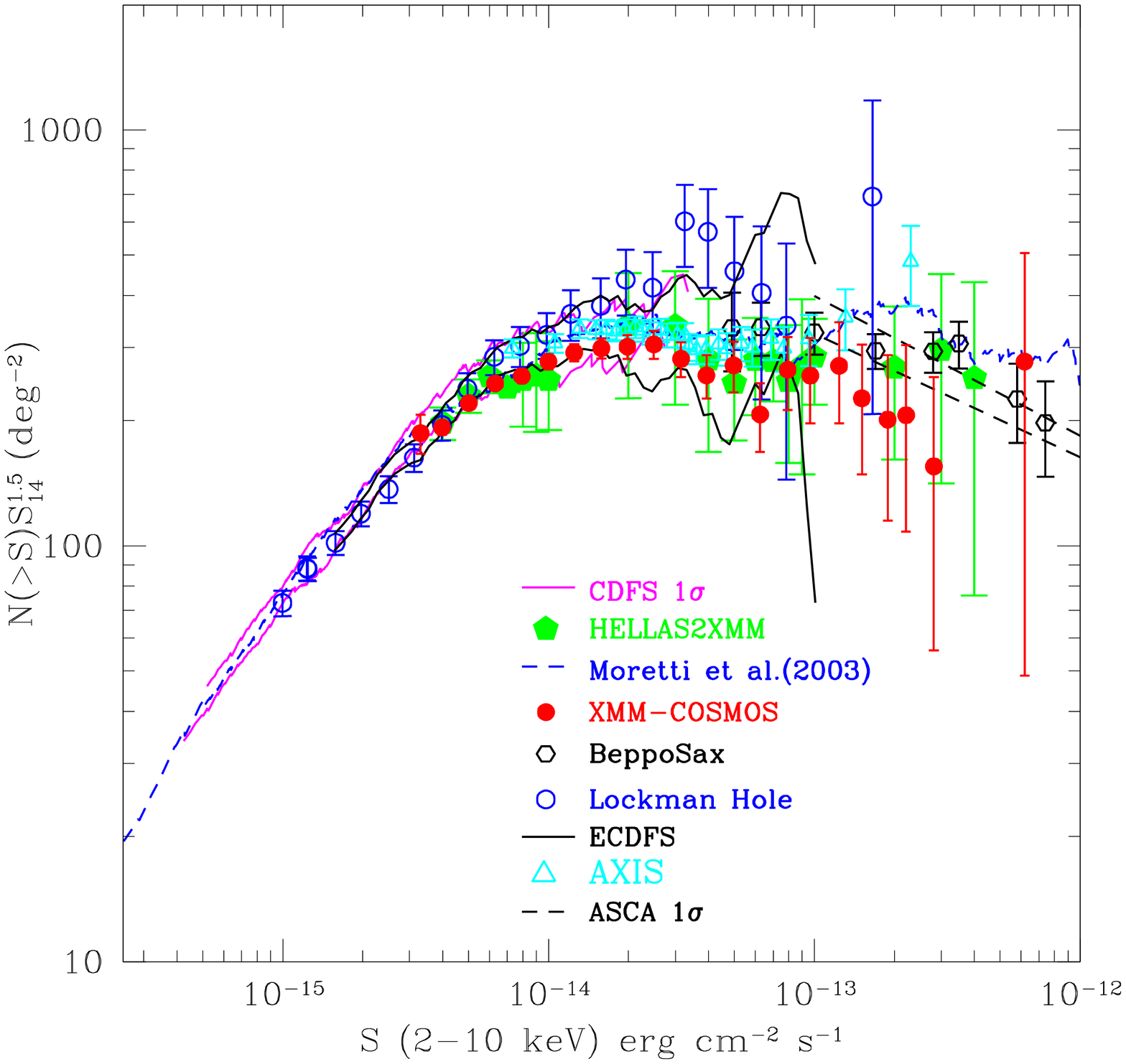}}
\caption{\label{fig:logn2}   The 2-10 keV logN-logS of  the XMM-COSMOS ($red~dots$) sources compared with
the combined {\it Chandra},
 \xmm~ and ASCA sources  \citep[][$blue~dashed~line$]{mor03}, the HELLAS BeppoSAX \cite[][$black~hexagons$]{gio00}
  the  2 Ms CDFS  \cite[1$\sigma$ error tie,][$magenta~continuous~line$]{luo}, 
 the  HELLAS2XMM  \cite[][$green~pentagons$]{bal}, the AXIS \cite[][$cyan~triangles$]{carrera07}, the extended CDFS 
  \citep[ 1$\sigma$ error tie,][$black~continuous~line$]{leh05} and  the Lockman
hole  \cite[][$blue~open~circles$]{brunner08} surveys.  The $black-dashed-line$ are the 1$\sigma$ confidence contours
of the best fit to the logN-logS of the ASCA data \citep{cagnoni98}. The
 source number counts are plotted multiplied by (S/10$^{14}$)$^{1.5}$ in order 
to highlight the  deviations from the Euclidean behavior. }
\end{center}
\end{figure}

\begin{figure}[!t]
\begin{center}
\resizebox{\hsize}{!}
{\includegraphics{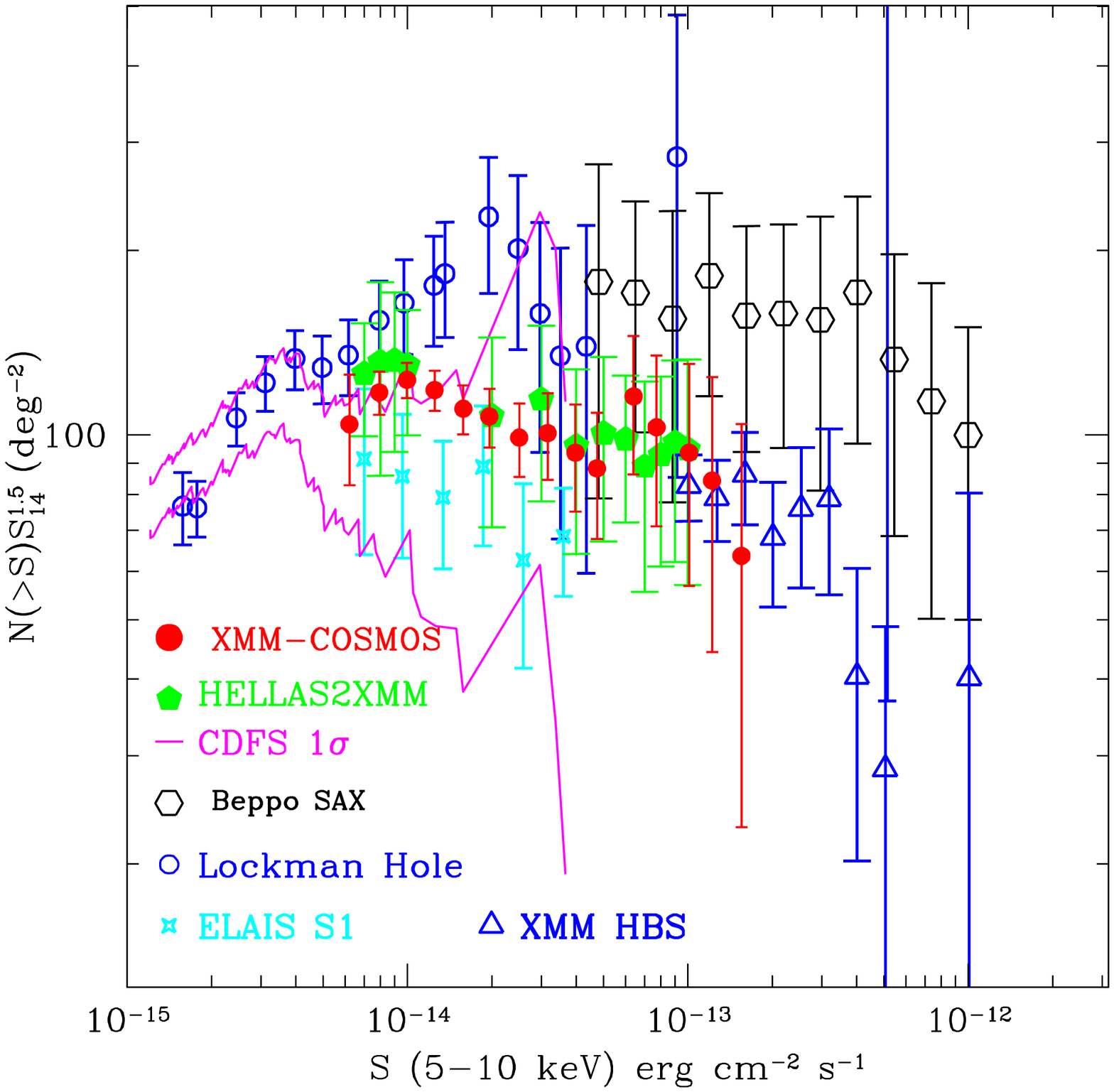}}
\caption{\label{fig:pippo}  The 5-10 keV logN-logS of the XMM-COSMOS ($red~dots$) sources compared with the 
HELLAS2XMM \citep[][$green~pentagons$]{bal}, the 1 Ms CDFS \citep[1$\sigma$ error tie,][$magenta~continuous~line$]{ros02}, 
the HELLAS-BeppoSAX  \citep[][$black~hexagons$]{fio01},  
the ELAIS S1  \citep[][$blue~stars$]{puc06}, the XMM-HBS  
\citep[][$blue~triangles$]{dc04} and the Lockman Hole  \citep[][$blue~open~circles$]{brunner08}. 
surveys. The source number counts are plotted multiplied by (S/10$^{14}$)$^{1.5}$ 
in order to highlight  deviations from the Euclidean behavior.   }
\end{center}
\end{figure}

Using the sky coverage we produced the cumulative logN-logS relations
 in the three energy bands under investigation by using:
\begin{equation}
N(>S)=\sum_{i=1}^{\it N_S} \frac{1}{\Omega_{i}} deg^{-2},
\end{equation}
where $N(>S)$ is the total number of detected sources in the field
with fluxes greater than $S$ and $\Omega_{i}$ is the sky coverage
associated with the flux of the i$^{th}$ source.  The variance of the source number counts is therefore
defined as:
\begin{equation}
\sigma_{i}^{2}=\sum_{i=1}^{\it N_S}\left(\frac{1}{\Omega_{i}}\right)^{2}.
\end{equation} 
The cumulative number counts, normalized to the Euclidean slope (i.e. multiplied by
S$^{1.5}$), are shown in 
Figure \ref{fig:logn1}, \ref{fig:logn2} and \ref{fig:pippo}, 
in the 0.5--2 keV, 2--10 keV and 5--10 keV energy ranges, respectively.
The logN-logS relations are also presented in Table \ref{tab:logn}.
From $left$ to $right$: Flux, Number-counts and area in the 
0.5--2 keV, 2--10 keV and 5--10 keV energy band, respectively.

In order to parametrize our relations, 
we performed a maximum likelihood fit  to the unbinned differential counts.
We assumed a broken power-law  model for the 0.5--2 keV and 2--10 keV bands:
\begin{equation}
n(S)=\frac{dN}{ds}= \left\{\begin{array}{ll}
A\,S^{-\alpha_{1}} & S>S_{b} \\
B\,S^{-\alpha_{2}} & S \leq S_{b},\\
\end{array}
\right.
\end{equation}

where $A=B\,S_{b}^{\alpha_{1}-\alpha_{2}}$ is the normalization, 
$\alpha_{1}$ is the bright end slope, $\alpha_{2}$ the faint end slope, 
$S_{b}$  the break flux, and $S$ the flux in units of 10$^{-14}$ \flux. 
 Notice that, using  the maximum likelihood method, the fit is not dependent on the  data binning
and therefore we are using the whole dataset. Moreover, the normalization $A$ is not 
a parameter of the fit, but  is obtained by imposing the condition that the number of expected 
sources from the best  fit model is equal to the  total observed number. \\
 In the  0.5--2 keV energy band the best fit parameters are $\alpha_{1}$=2.40$\pm{0.05}$, 
$\alpha_{2}$=1.60$^{+0.04}_{-0.10}$, $S_{b}$=1.00$^{+0.21}_{-0.26}\times$10$^{-14}$\flux~ and A=141. 
These values are consistent with those measured in 
paper II while the normalization is  lower than the value (A=198) derived in paper II\footnote{  Note 
that in paper II we gave the 
normalization of the cumulative distributions}. However, with  this  fitting method
the normalization  is not a fit parameter and it  
is strongly dependent on the best fit values of the bright end slope
and on the cut-off flux. One can indeed notice that the best fit values of the  $\alpha_{1}$ and
S$_{b}$ parameters are somewhat changed with respect to paper II.
The bright end slope varied from 2.6 in paper II to 2.4 in the present
work and the cut-off flux varied from $\sim$1.55$\times$10$^{-14}$\flux~to 1.00$\times$10$^{-14}$\flux. 
However, a comparison of the amplitude of the source surface density measured in paper II with that
measured here can be performed if we measure the model predicted source counts at
fluxes fainter than the knee. If we take 2$\times$10$^{-15}$ \flux~ as a reference flux,
in paper II we had a source surface density of  478 deg$^{-2}$ while  here we
measure  479 deg$^{-2}$. We can therefore conclude that the 0.5--2 keV 
logN-logS obtained in paper II  and in this work are  in good agreement.\\ 
In the  2--10 keV band the best fit parameters  are
 $\alpha_{1}$=2.46$\pm{0.08}$, $\alpha_{2}$=1.55$\pm{0.18}$, 
$S_{b}$=1.05$\pm{0.16}$ 10$^{-14}$\flux~ and A=413.
 Since the best fit parameters are similar to those of Paper II, we can directly
compare the normalizations of the logN-logS. The normalization derived in this work 
is 10\% higher than that measured in paper II.  This effect is partly due to 
 the sources missing in the 2--4.5 keV  band and detected in the 2--8 keV
which were not considered in the analysis of paper II. Moreover, extrapolation of 
the 2--4.5 keV count-rate into the broader 2--10 keV band
is  more affected by uncertainties on the true source spectral slope and provides wrong
 count-rate estimates especially for the most absorbed sources. \\  
In the 5--10 keV energy band we did not find  any  significant break in the slope. We therefore 
fitted the data using a single power-law in the form of n(S)=AS$^{-\alpha_{1}}$ and 
obtained  $\alpha_{1}$=2.38$\pm{0.05}$ and A=130.\\
 In the 5--10 keV both  the normalization and the slope are consistent within 1$\sigma$ with the values
obtained in paper II. 
\subsection{Comparison with previous surveys}
 In Fig. \ref{fig:logn1}, \ref{fig:logn2},\ref{fig:pippo} we compare our logN-logS
with  the results of previous surveys.   
A visual inspection of the data shows that the \cosm~source counts are
in general agreement, within $1\sigma$, with all the previous measurements. 
In the 0.5--2 keV band  source counts of all the surveys agree with our measurements,
with the only exception of the bright end of the Lockman Hole logN-logS.
The reason for such a discrepancy is that the location of the Lockman Hole survey was
chosen on purpose near a concentration of bright sources to improve
the accuracy of the ROSAT star tracker in order to achieve a 
better astrometry (G. Hasinger, private communication).
This had the result of artificially   increasing
the source counts at the bright end of the relation.
The comparison with   other surveys is consistent with the error bars
and with the counts in cell fluctuations predicted in Paper II. 
We also compared our results with the recent work of \citet{mateos08} who performed 
a detailed analysis of the logN-logS of X-ray sources detected in 1129 \xmm~archival
observations. By comparing the  data of Table \ref{tab:logn} with those shown in 
Table 3 of \citet{mateos08} we found 
1$\sigma$ agreement in almost all the data bins.   \\
In paper II we showed that the fluctuations of the source counts 
are proportional to the actual number of sources in the field and to the amplitude of the
angular auto-correlation function of the X-ray sources. Therefore, assuming a universal
shape of the autocorrelation function,
we expect that the surveys showing the largest deviations from the mean value
of the source density are the pencil beam surveys (i.e. area $<$0.2 deg$^{2}$) at their bright end.
Moreover, 
with \cosm~, fluctuations introduced in previous shallow surveys by low counting statistics   and
by random sampling of a few large structures in pencil beam surveys are largely suppressed.
With the same formulas used  in Paper II, 
we estimate the effect of the cosmic variance to be $<$5\% on the normalization 
of the \cosm~logN-logS and that the new data do not change the results shown in Paper II. \\
 Also in the 2--10 keV energy bands we do not  note any significant deviation from 
previous works with the exception only of the Lockman Hole and  the two faintest bins of
the AXIS counts.  Also in this band our data are in good agreement with the results of 
\citet{mateos08}.  \\
Fig. \ref{fig:pippo} suggests that the fluctuations of the source counts in the 5--10 keV band are much larger
than  in the other bands. This is due to the fact that as discussed above and in paper II, 
when we deal with low source surface density, the impact of the sample variance
becomes significantly high. 
However in this band  our data are statistically consistent with most of the data from other surveys.
Also in this energy band, the deviation of the COSMOS data from those of the Lockman Hole
is due the higher number of bright sources in that particular field. 
Our source counts are  10-15 \% higher than those of the ELAIS-S1 survey \citep{puc06}. 
As an example, using Eq. 10 of paper II, we determine
that at the faint end of the 5-10 keV band,  fluctuations due to the cosmic variance are of the 
order of the 20-40 percent, depending on the survey size. At the bright end large
deviations of more than a factor of two are still allowed by the sample variance. This is also visible
in Fig. \ref{fig:pippo} where at the bright end the Beppo-SAX counts exceed the \cosm~counts by about
a factor of two though remaining statistically consistent with each other.\\
\citet{kim07} reported the results of a broken power law fit to the logN-logS
from different surveys available in the literature.
They also reported measurements of the CHAMP survey 
which is a compilation of Chandra archival data  for a total sky coverage of 9.6 deg$^{2}$,
with  a depth about one order of magnitude fainter than \cosm.
On average the bright end slopes are consistent with a Euclidean rise in all the surveys. 
The faint end slopes are of the order of $\alpha_{2}\sim$1.5-1.6 in the 0.5-2 keV band and
span from $\alpha_{2}\sim$1.3 to $\alpha_{2}\sim$2.0 with a mean of $\alpha_{2}\sim$1.6-1.7.
A larger spread is reported for the  cut off fluxes. 
Although the spread in this  parameter is quite large, our data are consistent
 with the average values reported in the literature  for this parameter.  \\
\subsection{Extragalactic X-ray source number counts and comparison with models}
\begin{figure}[!h]
\begin{center}
\resizebox{\hsize}{!}
{\includegraphics{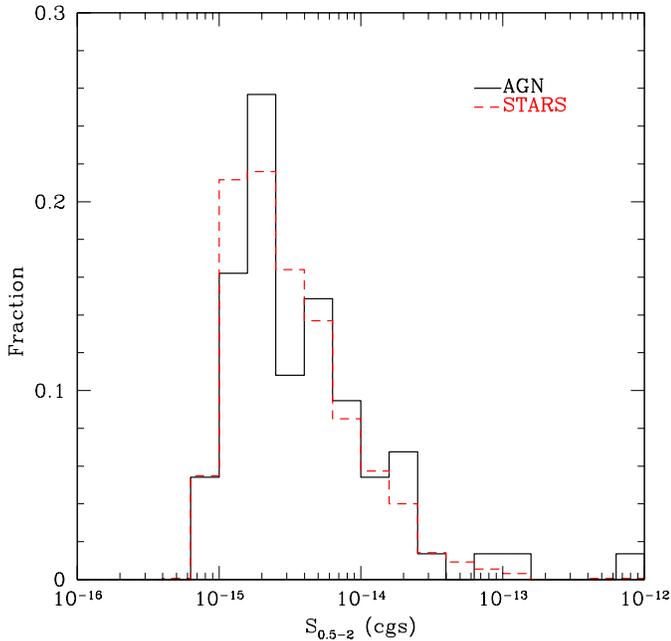}}
\caption{\label{fig:histstars} The 0.5--2 keV flux distribution of sources classified 
as AGN or extragalactic ($black$)  and stars ($red$). }
\end{center}
\end{figure}
\begin{figure}[!h]
\begin{center}
\includegraphics[scale=0.45]{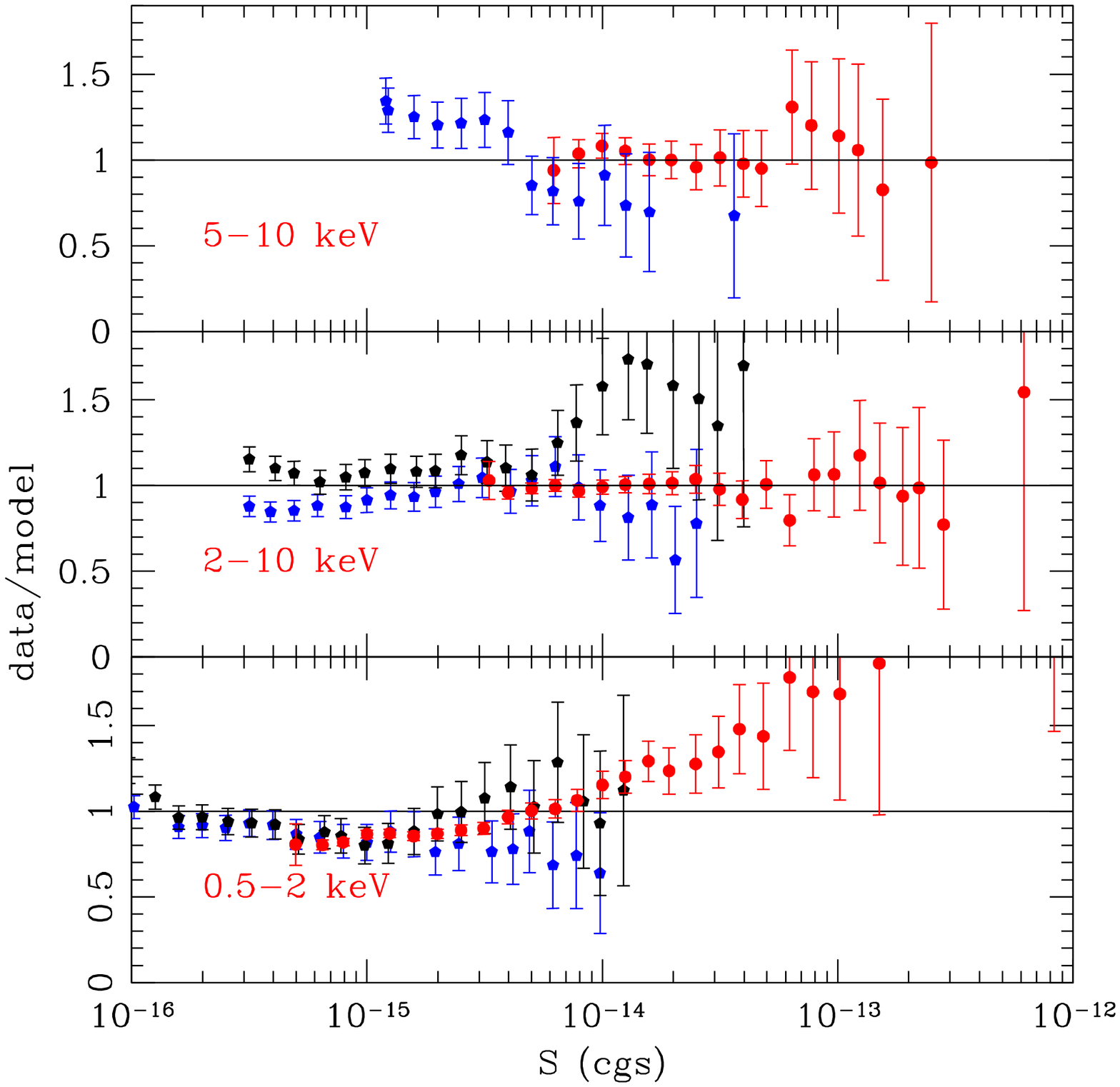}
\includegraphics[scale=0.45]{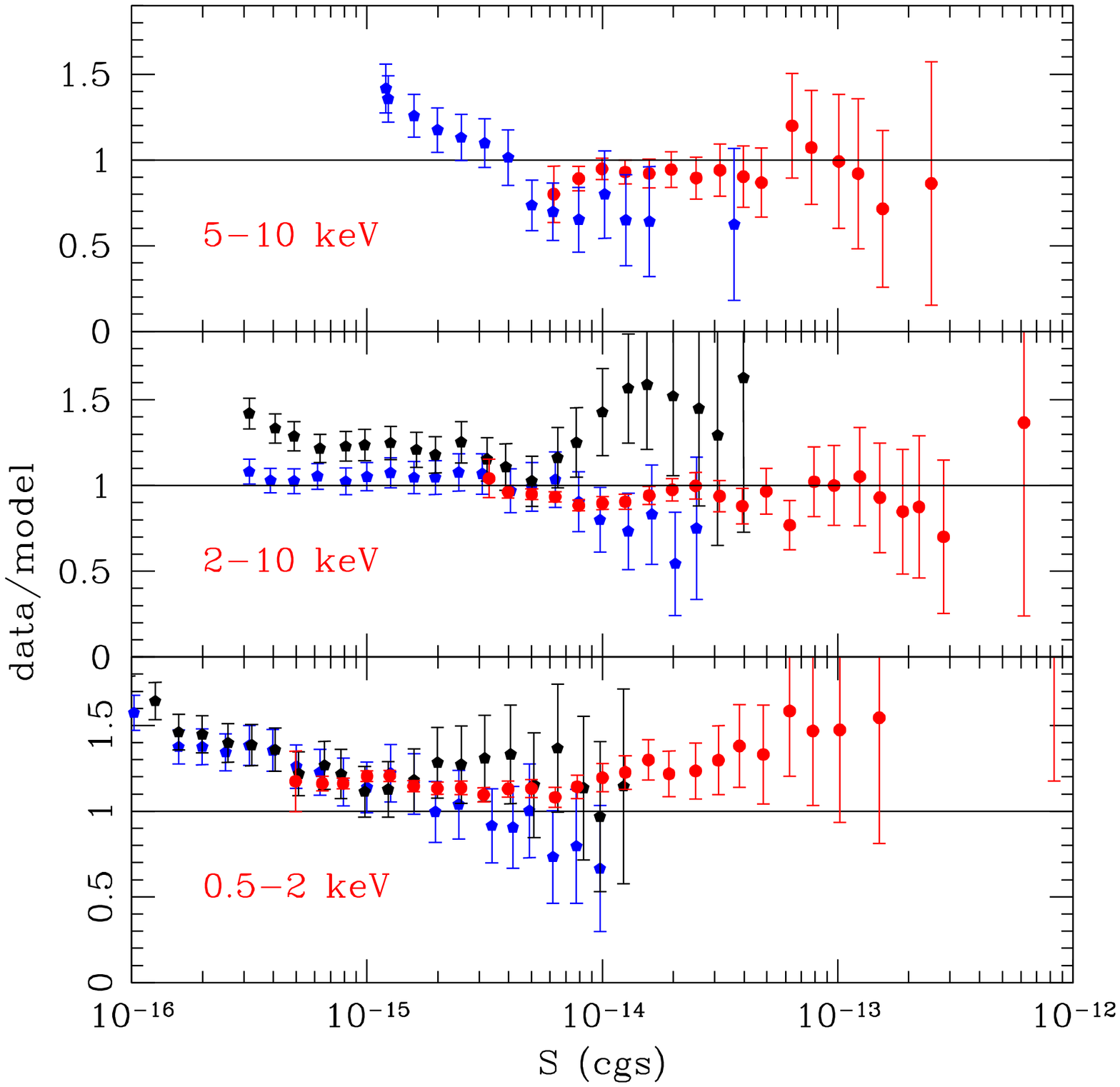}
\caption{\label{fig:datamod} $Upper~panel$ The ratio between the \citet{gilli07} model logN-logS relations 
to the observed source counts in \cosm~and in the Chandra deep fields \citep{ros02,bau04}. From bottom to top
in the 0.5--2 keV, 2--10 keV and 5--10 keV energy bands. The \cosm~datapoints are plotted in red, while in black and blue we plot
the CDFN and CDFS, respectively. $Bottom~panel$: Same as $Upper~panel$ but using  the \citet{treister06} XRB model.}
\end{center}
\end{figure}
We  used our logN-logS relations  to test the most recent  extragalactic XRB synthesis models.
 In order to compare our data with the XRB model, we estimate the fraction of sources classified as stars by \cite{brusa09}.
In the 0.5-2 keV band we identified 74/1621 (i.e. $\sim$4.5$\%$) sources classified as stars, while
 these are 17/1111 (i.e. $\sim$1.5$\%$) and  3/251 (i.e. $\sim$1.1$\%$)
in the 2-10 keV and 5--10 keV bands, respectively.
In figure \ref{fig:histstars} we plot the normalized distributions of the fluxes of stars and extragalactic
sources in the 0.5-2 keV band. Since the  two distributions are similar we  can conclude that stars 
in the \cosm~flux range  affect the 0.5-2 keV logN-logS only by 
increasing the extragalactic source counts by $\sim$5\%. 
\citet{mateos08} measured a flux dependent fraction of stars, with higher 
fractions than ours at bright fluxes where \cosm~is undersampled.  
By excluding the source classified as stars, we derived the logN-logS
relations for extragalactic sources only.\\
In the upper panel of Fig. \ref{fig:datamod} we plot the ratio of the \cosm~logN-logS 
relations to the predictions of the XRB population synthesis model of \citet{gilli07} (hereafter model I) ,
 while in the bottom panel we plot for comparison the ratio of the data to
 the model of \citet{treister06}  (hereafter model II). 
 In both  models the XRB spectrum is dominated by obscured AGN 
 which outnumber unobscured ones by a factor 3--4 at low X--ray luminosities
 (log$L_X <$ 44). The cosmological evolution is similar and parametrized using 
 the most recent determinations of the AGN luminosity function 
 (Ueda et al. 2003; La Franca et al. 2005; Hasinger et al. 2005).
 In both  models the obscured fraction decreases towards high luminosity.
 The luminosity dependence is stronger in Treister et al. (2006)
 who also allow the obscured fraction to increase at high redshifts.
 The absorption distribution is peaked around log$N_{H} \sim$ 23.5  in Gilli 
 et al. (2007), while it remains rather flat above log$N_{H} \sim$ 22  in Treister 
 et al. (2006). 
 They also differ in the adopted  XRB intensity around the 30 keV peak. 
 The Gilli et al. (2007) model is tuned to fit the HEAO-1 level, consistent, 
 within 10\%, with recent BeppoSAX (Frontera et al. 2007) and Swift BAT 
 (Ajello et al. 2008) measurements, while Treister et al. renormalize 
  the HEAO-1 intensity upward by a factor 1.4 to better match the extrapolation 
 of lower energy ($<$ 10 keV) data (i.e. De Luca \& Molendi 2004). 
Moreover, for  this paper
we adopt a modified version of the \citet{gilli07} model\footnote{The predictions  of the model can be retrieved on line
at {\bf http://www.oabo.inaf.it/$\sim$gilli/counts.html}
 using the POMPA COUNTS software (POrtable Multi Purpose Application for the AGN COUNTS). } 
 which takes into account the decline
of the space density of AGN at z$>$3  discussed  by \citet{brusa08}.
In order to test the models over a wider range of fluxes we also plotted the data of the
CDFN \citep{bau04} and CDFS \citep{luo} surveys. By restricting our analysis to  fluxes larger than 10$^{-16}$ \flux,
 the contribution of normal galaxy counts is negligible \citep{ranalli}.
The results of this comparison can be   summarized as follows:
\begin{itemize} 

\item 	In the 5--10 keV energy band, both  models reproduce well  the \cosm~logN-logS, while the CDFS counts
	show a systematically  different slope  from that of the predicted relation. However, because of
	the small effective area of Chandra above 5 keV (i.e. $\sim$200 cm$^{2}$ at 6.4 keV), 
	the 5--10 keV CDFS logN-logS may suffer from significant systematic uncertainties . 

\item   In the 2--10 keV energy band, the models reproduce quite accurately the
	\cosm~data, although  model II slightly (i.e. $\sim$ 10\%) overpredicts
	the \cosm~counts. The CDFN counts show a systematically higher normalization 
	than those of the models (up to 40\% at faint fluxes) and of the COSMOS and CDFS data.

\item   In the 0.5--2 keV band both  models show significant deviations from both  data sets. 
 	Source counts estimated from model I show a systematically  steeper slope than the data.
	 On average, model I deviates from the observations by about 10-15\% in the flux range
	10$^{-16}$ \flux-10$^{-13}$ \flux. Model II, on the other hand, systematically 
	underestimates the source counts while a visual inspection shows a good 
	agreement with the observed slopes.  In this case deviations
 	 between the data and the model are of the order of  20\% at the \cosm~fluxes, while at fainter
	fluxes the deviations are larger and of the order of 30\%-40\%. Both models underestimate the observed 
	counts by $\sim$30\% at fluxes greater than $\sim$10$^{-14}$ \flux
\end{itemize}
In summary, the hard X-ray observations are very well 
reproduced by both  models with an accuracy of 
$\sim$10\%. In the soft band the agreement between the predicted and the
observed relations is not as good as in the harder energy bands\footnote{Note
 that the inclusion of a decline in the space density 
of AGN at high-z in model I affects mostly the 0.5--2 keV
energy bands. In the harder bands the predicted number counts
are comparable with or without a high-z space density decline.} .
 The level of the discrepancy, however, is  small  ($\sim$20\%) and such that
it can be easily accommodated by slight variations of the XRB model parameters.
 This band is in fact more sensitive to the effect of absorption and therefore a fine 
tuning of absorption in AGN is required in the models. Moreover, this band contains a larger fraction
of high-z objects \citep{brusa09}.
Therefore, the fact that the two models assume a somewhat  different
absorption evolution and XRB spectrum can, in the first instance,
 explain the different source count predictions. 
We can conclude that at the flux limits of the \cosm~survey,
XRB synthesis models can reproduce the observations with a precision of 10\%-20\%. 
\section{\label{xrc} X-ray colours of the X-ray sources}

\begin{figure}[!t]
\begin{center}
\resizebox{\hsize}{!}
{\includegraphics{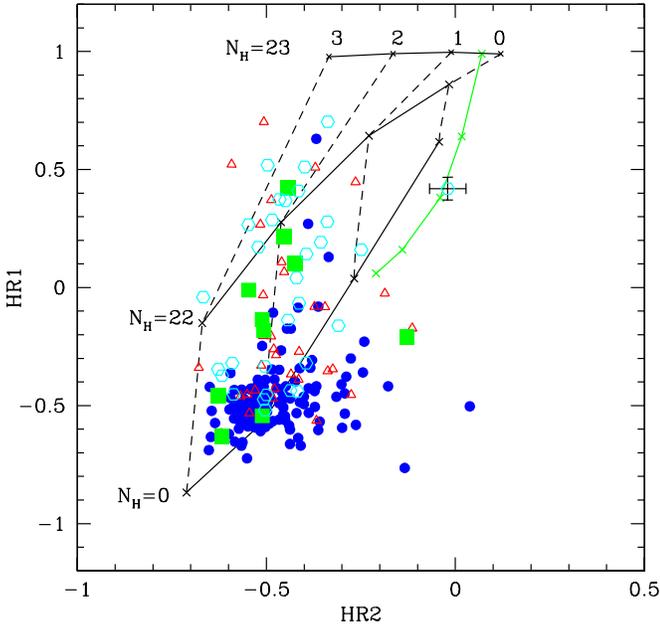}}
\caption{\label{fig:hr}   
	X-ray colour-colour diagram in the \cosm~survey. Colours are defined in the text. 
	 XID \#2608  has been plotted with its error which
	also represents the typical amplitude of the uncertainties in the plot. 
		The grid represents the places in the  HR1-HR2 plane of sources with 
		single power-law spectra with $\Gamma$=0$\div$3 with absorption with a column density  
		log(N$_{H}$)=0$\div$23 cm$^{-2}$. The green line marks the region occupied by candidate Compton thick-AGN
		[i.e. log(N$_{H}>$24) cm$^{-2}$] and the marks on top of it represent  1\%, 3\% 10\% and 30\% level 
		 of leaking flux, from top to bottom.   We represent Type I AGN, Type II AGN, emission line and absorption 
		line galaxies as $blue~filled~circles$, $red~empty~triangles$, $cyan~empty~exagons$, $green~filled~squares$, respectively.   }
\end{center}
\end{figure}

\begin{figure}[!t]
\begin{center}
\resizebox{\hsize}{!}
{\includegraphics{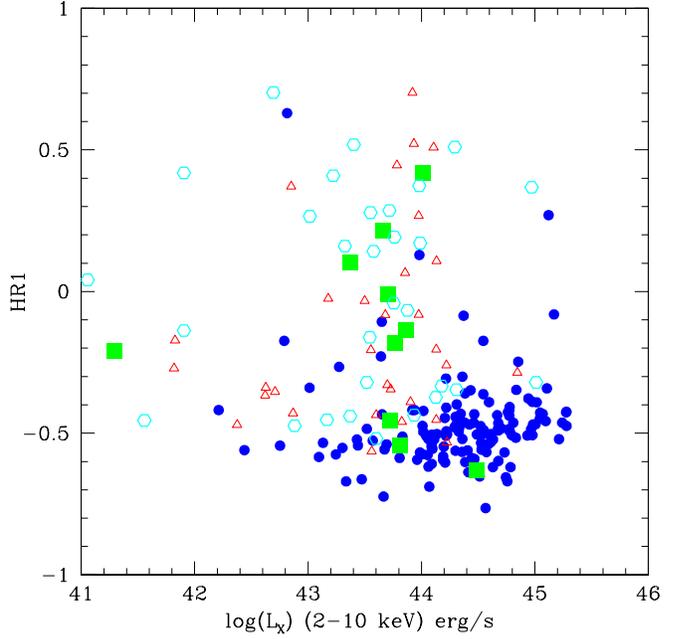}}
\caption{\label{fig:lumhr}  The 2--10 keV X-ray luminosity vs. HR1 for the sources fullfilling the 
HR error selection.  We represent Type I AGN, Type II AGN, emission line and absorption 
		line galaxies as $blue~filled~circles$, $red~empty~triangles$, $cyan~empty~exagons$, $green~filled~squares$, respectively.  } . 

\end{center}
\end{figure}

The X-ray colours or hardness ratios are defined as  \
\begin{equation}
 HR_1 = \frac{B_2 - B_1}{B_2 + B_1} ~and~HR_2 = \frac{B_3 - B_2}{B_3 + B_2}
\end{equation}
where $B_1$, $B_2$, and $B_3$ refer to the vignetting-corrected count rates 
in the 0.5--2 keV, 2--10 keV and  5--10 keV bands, respectively. By construction,
both HR1 and HR2 can assume values between -1 and 1.\\
Fig. \ref{fig:hr} displays  the HR1-HR2 plot of 212 sources
for which the 1$\sigma$ error on both HR1 and HR2 is $<$0.25 and for
which a high quality optical spectrum is available.
The plot also contains  a grid of the  expected values of HR1 and HR2 for different spectral models.
In particular we considered a simple power law model with a  spectral index in the interval $\Gamma$=0$\div$3 
and  with   a column density  log(N$_{H}$)=0$\div$23 cm$^{-2}$.\\
In \citet{brusa09}, and \citet{trump}, extragalactic sources are  classified into 4
main categories:
\begin{itemize}
\item Type I AGN, if the optical spectrum shows evidence of broad (FWHM$>2000$ Km s$^{-1}$) emission lines;
\item Type II AGN, if the optical spectrum shows evidence of   narrow, high-ionization emission lines and/or AGN diagnostic diagrams;
\item Emission line galaxy, if the optical spectrum is dominated 
	by a galaxy continuum plus emission lines but without
	secure  AGN indicators;
\item Absorption line galaxy if the optical spectrum is dominated 
	by a galaxy continuum plus absorption lines;
\end{itemize}
Details of the optical classification of X-ray sources are extensively discussed in 
\citet{brusa09}, therefore we limit our analysis to the X-ray properties of these sources.
140/212 are classified as Type I, 32/212 as Type II,  30/212  
as emission line galaxies and 10/212 as absorption line galaxies.
Note that with the exception  only of 7 objects, 
all the  sources have an estimated 
2--10 keV X-ray luminosity log(L$_{X}$)$>$42 erg/s, with most of them
having log(L$_{X}$)$>$43 erg/s (see Fig. \ref{fig:lumhr}).
The adopted cuts on the errors on the HR  preferentially select
unabsorbed to moderate absorbed AGN, biasing the sample against 
normal galaxies, starforming galaxies
and the most obscured AGN.  \\
Type I AGN  ($blue~filled~circles$) cluster in   a region around HR1=-0.5 and HR2=-0.5 with a relatively small dispersion,
corresponding to a typical  X-ray spectrum dominated by a power-law continuum with very low absorption.
 Only a few  Type I sources have X-ray colours typical of Type II sources (i.e. HR$>-0.1$ which corresponds to N$_{H}>$10$^{22}$ cm$^{-2}$).
This fraction ($\sim$2\%)  is consistent with the results from X-ray spectral analysis on a subsample of \cosm~sources
 (Mainieri et al. 2007) but at variance with previous works on the fraction of X-ray absorbed Type I AGN 
at comparable X-ray luminosity (see e.g. Brusa et al. 2003, Perola et al. 2004, Page et al. 2004) which reported values
as large as 10\%.  However such a low fraction may be a consequence of the selection effect mentioned above.\\
On the other hand, type II AGN ($red~empty~triangles$)  fill most of the HR1-range, corresponding
to observed frame absorption up to 10$^{23}$ cm$^{−2}$.
The fraction of Type II AGN with X--ray colours typical of Type I AGN (HR1$<-0.3$) is $\sim$30\%. 
This is consistent with the fraction of X--ray unobscured Type II AGN reported in 
Mainieri et al. (2007). \\
An interesting source is 
  XID=\#2608 which has been classified as a
Compton-Thick AGN by \citet{mainieri07} and \citet{has07}
but its optical spectrum is that of  an
emission line galaxy.   
In \citet{has07} a small number of sources (including XID=\#2608) 
was found to have hardness ratios that
could be interpreted as being due to heavily absorbed (possibly Compton thick)
high energy spectra with some fraction of leaking 
unabsorbed soft flux. The solid green  line
in Fig. \ref{fig:hr} represents the expected 
tracks occupied by leaking Compton thick sources
in the HR1-HR2 plane\footnote{The position in this HR1-HR2 plane of the track of 
leaking Compton thick objects is different from 
that shown in \citet{has07} because of the 
difference in the energy range of the hard band 
(i.e. 2--8 keV here, 2--4.5 keV in \citet{has07}).} at z=0.
The line has been computed with a pure reflection  model with a fraction 
of  1\%,3\%,10\% and 30\% (from $top$ to $bottom$) of 
the flux from the central source leaking out.\\
In particular the source ID=\#2608 shows 
X-ray colours typical of a spectrum dominated by 
a pure reflection component with $\sim$3\% of the original flux leaking out. 
Another source,  XID=\#131, shows X-ray colours consistent with Compton-thick AGN with a 
small fraction of leaking flux. 
We note that a 1\% fraction of Compton-thick AGN is 
consistent with the predictions of XRB models at the
flux limit of this subsample (i.e. 2--10 keV flux $>$10$^{-14}$ \flux)
and with the source counts of Compton-thick objects measured in a collection
of surveys  by \citet{brunner08}.\\
 The objects classified as emission and absorption line galaxies
are spread over the entire luminosity-hardness ratio plane
(see Fig. \ref{fig:lumhr}) and their nature can be explained as a mixture
of star forming galaxies, Type II AGN and XBONGs \citep[see e.g.][]{comastri,caccianiga,civano,cocchia}. 
A more detailed
analysis of their multiwavelength properties will be the subject of
a forthcoming publication.

\section{\label{sum} Summary} 
In this paper we presented a pointlike source catalogue in the \cosm~survey. The survey
covers an area of 2.13 deg$^{2}$ in the equatorial sky.
The field has been observed with 55 \xmm~pointings
for a total exposure time of $\sim$1.5 Ms. We achieved an almost uniform
exposure of  $\sim$40 ks on the field. \\
We detected a total number of 1621, 1111 and 251 sources in the 0.5--2 keV, 
2--10 keV and 5--10 keV energy band, respectively, for a total of 
1887 independent sources detected with det\_ml$>$10 in at least  one band. 
 The survey has a limiting flux  of $\sim$1.7$\times$10$^{-15}$ \flux,
 $\sim$9.3$\times$10$^{-15}$ \flux~ and $\sim$1.3$\times$10$^{-14}$ \flux~ 
in the 0.5--2 keV, 2--10 keV and 5--10 keV energy band, 
over 90\% of the area.  \\
 Together with the source catalogue we derived logN-logS relations with high statistics  
in the flux interval sampled by the survey. The logN-logS relations
are in good agreement with most of the X-ray surveys published in the
literature. We compared our source counts with the most recent XRB population synthesis
models \citep{gilli07,treister06} and found that they agree within 10\% with our data
in the 5--10 keV and 2--10 keV energy bands. In the 0.5--2 keV band
both  models deviate from the \cosm~data by about 10\%-30\% suggesting, that further
improvements in the modeling are required.
 We isolated a subsample of X-ray bright sources for which optical spectroscopy is
available. About 65\% of them have optical and X-ray properties typical
 of   Type I AGN and  $\sim$15\% of Type II AGN.  
In the subsample of sources with a good  optical spectrum and good counting 
statistics, the number of candidate Compton thick (1-2)
AGN is fully consistent with the expectations of XRB population synthesis models.
By combining X-ray colours and optical spectroscopy   
we  found that 20\% of the sources do not show,
 in the optical band, evident signatures  of AGN activity although their
 X-ray luminosities are typical of AGN. 
Additonally, we consider \cosm~as a pathfinder for the
eROSITA  \citep{rosita} X-ray telescope which will be launched in 2012 and that
will perform an all sky survey with sensitivities comparable to those presented here.
\begin{acknowledgements}
      Part of this work was supported by the German
      \emph{Deut\-sche For\-schungs\-ge\-mein\-schaft, DFG\/} project
      number Ts~17/2--1.In Germany the \xmm~project is supported by 
	the Bundesministerium f\"ur Wirtshaft und Techologie/Deutsches
	Zentrum f\"ur Luft und Raumfahrt and the Max-Planck society.
	NC and AF were partially supported from a
	 NASA grant NNX07AV03G to UMBC. In Italy, the \cosm~project
	is supported by PRIN/MIUR under grant 2006-02-5203 and ASI-INAF grants I/023/05/00, I/088/06.
	 Part of this work was supported by the German Deutsche
	Forschungsgemeinschaft,DFG Leibniz Prize (FKZ HA 1850/28-1).
	NC gratefully acknowledges
	Ezequiel Treister for providing his prediction of the XRB logN-logS.
	The entire COSMOS collaboration is gratefully acknowledged.
\end{acknowledgements}


\begin{thebibliography}{}
\bibitem[Ajello et al.(2008)]{ajello} Ajello, M., et al.\ 
2008, arXiv:0808.3377 
\bibitem[Baldi et al.(2002)]{bal} Baldi, A., Molendi, S.,
Comastri, A., Fiore, F., Matt, G., \& Vignali, C.\ 2002, \apj, 564, 190
\bibitem[Bauer et al.(2004)]{bau04} Bauer, F.~E., Alexander,
D.~M., Brandt, W.~N., Schneider, D.~P., Treister, E., Hornschemeier, A.~E.,
\& Garmire, G.~P.\ 2004, \aj, 128, 2048
\bibitem[Branchesi et 
al.(2007)]{bra07} Branchesi, M., Gioia, I.~M., Fanti, C., Fanti, R., \& Cappelluti, N.\ 2007, \aap, 462, 449 
\bibitem[Brunner et al.(2008)]{brunner08} Brunner, H., Cappelluti, N., Hasinger, G., Barcons, X., Fabian, A.~C., Mainieri, V., \& Szokoly, G.\ 2008, \aap, 479, 283 
\bibitem[Brandt 
\& Hasinger(2005)]{brandt} Brandt, W.~N., \& Hasinger, G.\ 2005, \araa, 43, 827 
\bibitem[Brusa et al.(2008)]{brusa08} Brusa, M., et al.\ 2008, 
arXiv:0809.2513 
\bibitem[Brusa et al.(2009)]{brusa09} Brusa, M. \etal~  2009, \aap, in preparation
\bibitem[Caccianiga et al.(2007)]{caccianiga} Caccianiga, A., Severgnini, P., Della Ceca, 
R., Maccacaro, T., Carrera, F.~J., \& Page, M.~J.\ 2007, \aap, 470, 557 
\bibitem[Cagnoni et al.(1998)]{cagnoni98} Cagnoni, I., della 
Ceca, R., \& Maccacaro, T.\ 1998, \apj, 493, 54 
\bibitem[Cappelluti et 
al.(2005)]{cap05} Cappelluti, N., Cappi, M., Dadina, M., Malaguti, G., Branchesi, M., D'Elia, V., \& Palumbo, G.~G.~C.\ 2005, \aap, 430, 39 
\bibitem[Cappelluti et al.(2007a)]{cap07} Cappelluti, N., et 
al.\ 2007, \apjs, 172, 341 
\bibitem[Cappelluti et 
al.(2007b)]{cap07a} Cappelluti, N., B{\"o}hringer, H., Schuecker, P., Pierpaoli, E., Mullis, C.~R., Gioia, I.~M., \& Henry, J.~P.\ 2007, \aap, 465, 35 
\bibitem[Cappi et al.(2001)]{cappi01} Cappi, M., et al.\ 2001, 
\apj, 548, 624 
\bibitem[Carrera et al.(2007)]{carrera07} Carrera, F.~J., et al.\ 2007, \aap, 469, 27 
\bibitem[Cavaliere \& Fusco-Femiano(1976)]{cav} Cavaliere,
A., \& Fusco-Femiano, R.\ 1976, \aap, 49, 137
\bibitem[Civano et 
al.(2007)]{civano} Civano, F., et al.\ 2007, \aap, 476, 1223 
\bibitem[Cocchia et 
al.(2007)]{cocchia} Cocchia, F., et al.\ 2007, \aap, 466, 31 
\bibitem[Comastri et al.(2002)]{comastri} Comastri, A., et al.\ 
2002, \apj, 571, 771 
\bibitem[Della Ceca et al.(2004)]{dc04} Della Ceca, R., et
al.\ 2004, \aap, 428, 383
\bibitem[De Luca 
\& Molendi(2004)]{2004A&A...419..837D} De Luca, A., \& Molendi, S.\ 2004, \aap, 419, 837 
\bibitem[Elvis et al.(2009)]{elvis08} Elvis, M., et
al.\ 2009, \apj, submitted
\bibitem[Finoguenov et al.(2007)]{fino07} Finoguenov, A., et 
al.\ 2007, \apjs, 172, 182 
\bibitem[Finoguenov et al.(2008)]{fino08} Finoguenov, A., et 
al.\ 2008, \apj in preparation
\bibitem[Fiore et al.(2001)]{fio01} Fiore, F., et al.\ 2001,
\mnras, 327, 771
\bibitem[Fiore et al.(2003)]{fio03} Fiore, F., et al.\ 2003,
\aap, 409, 79
\bibitem[Frontera et al.(2007)]{2007ApJ...666...86F} Frontera, F., et al.\ 
2007, \apj, 666, 86 
\bibitem[Gilli et al.(2007)]{gilli07} Gilli, R., Comastri, A., \& Hasinger, G.\ 2007, \aap, 463, 79 
\bibitem[Giommi et al.(2000)]{gio00} Giommi, P., Perri, M.,
\& Fiore, F.\ 2000, \aap, 362, 799
\bibitem[Hasinger et al.(1993)]{has93} Hasinger, G., Burg, R., Giacconi, R., Hartner, G., Schmidt, M., 
Trumper, J., \& Zamorani, G.\ 1993, \aap, 275, 1 
\bibitem[Hasinger et al.(2005)]{has05} Hasinger, G., Miyaji,
T., \& Schmidt, M.\ 2005, \aap, 441, 417
\bibitem[Hasinger et al.(2007)]{has07} Hasinger, G., et al.\ 
2007, \apjs, 172, 29 
\bibitem[Hathaway et al.(1999)]{hat99} Hathaway, D.~H., 
Wilson, R.~M., \& Reichmann, E.~J.\ 1999, \jgr, 104, 22375 
\bibitem[Kenter et al.(2005)]{ken05} Kenter, A., et al.\
2005, \apjs, 161, 9
\bibitem[Kim et al.(2007)]{kim07} Kim, M., et al.\ 2007, 
\apjs, 169, 401 
\bibitem[Kocevski et al.(2008)]{koc08} Kocevski, D.~D., 
Lubin, L.~M., Gal, R., Lemaux, B.~C., Fassnacht, C.~D., 
\& Squires, G.~K.\ 2008, ArXiv e-prints, 804, arXiv:0804.1955 
\bibitem[La Franca et al.(2005)]{2005ApJ...635..864L} La Franca, F., et 
al.\ 2005, \apj, 635, 864 
\bibitem[Lehmer et al.(2005)]{leh05} Lehmer, B.~D., et al.\
2005, \apjs, 161, 21
\bibitem[Luo et al.(2008)]{luo} Luo, B., et al.\ 2008, 
ArXiv e-prints, 806, arXiv:0806.3968 
\bibitem[Mainieri et al.(2007)]{mainieri07} Mainieri, V., et al.\ 
2007, \apjs, 172, 368 
\bibitem[Mainieri et al.(2008)]{mainieri08} Mainieri, V., et al.\ 
2009, \aap, in preparation 
\bibitem[Mateos et al.(2008)]{mateos08} Mateos, S., et al.\ 
2008, arXiv:0809.1939 
\bibitem[McCracken et al.(2007)]{mac07} McCracken, H.~J., et 
al.\ 2007, \apjs, 172, 314 
\bibitem[Moretti et al.(2003)]{mor03} Moretti, A., Campana,
S., Lazzati, D., \& Tagliaferri, G.\ 2003, \apj, 588, 696
\bibitem[Narsky(2000)]{nar06} Narsky, I.\ 2000, Nuclear 
Instruments and Methods in Physics Research A, 450, 444 
\bibitem[Page et al.(2004)]{page} Page, M.~J., Stevens, 
J.~A., Ivison, R.~J., \& Carrera, F.~J.\ 2004, \apjl, 611, L85 
\bibitem[Panessa 
\& Bassani(2002)]{panessa} Panessa, F., \& Bassani, L.\ 2002, \aap, 394, 435 
\bibitem[Perola et 
al.(2004)]{perola} Perola, G.~C., et al.\ 2004, \aap, 421, 491 
\bibitem[Predehl et al.(2006)]{rosita} Predehl, P., et al.\ 
2006, \procspie, 6266,  
\bibitem[Puccetti et al.(2006)]{puc06} Puccetti  et al.\
2006, \aap, 457, 501
\bibitem[Ranalli et al.(2003)]{ranalli} Ranalli, P., Comastri, A., \& Setti, G.\ 2003, \aap, 399, 39 
\bibitem[Rosati et al.(2002)]{ros02} Rosati, P., et al.\
2002, \apj, 566, 667
\bibitem[Salvato et al.(2008)]{salvato08} Salvato, M., et al.\ 
2008, arXiv:0809.2098 
\bibitem[Scoville et al.(2007)]{scoville} Scoville, N., et al.\ 
2007, \apjs, 172, 1 
\bibitem[Schmitt \& Maccacaro(1986)]{sch} Schmitt,
J.~H.~M.~M., \& Maccacaro, T.\ 1986, \apj, 310, 334
\bibitem[Szokoly et al.(2004)]{zoccoli} Szokoly, G.~P., et al.\ 
2004, \apjs, 155, 271 
\bibitem[Treister 
\& Urry(2006)]{treister06} Treister, E., \& Urry, C.~M.\ 2006, \apjl, 652, L79 
\bibitem[Trump et al.(2009)]{trump} Trump, J.~R., et
al.\ 2009, \apj, accepted
\bibitem[Ueda et al.(2003)]{2003ApJ...598..886U} Ueda, Y., Akiyama, M., 
Ohta, K., \& Miyaji, T.\ 2003, \apj, 598, 886 





\end{thebibliography}
\end{document}